\documentclass[letterpaper, numberedappendix]{emulateapj} 

\usepackage{graphicx}
\usepackage{epsfig}
\usepackage{amsmath}
\usepackage{subfigure} 
\usepackage{hyperref}
\usepackage{amssymb}
\usepackage{lscape} 
\usepackage{apjfonts}
\usepackage{latexsym}
\usepackage[usenames,dvipsnames]{color}

\DeclareMathAlphabet{\mathitbf}{OML}{cmm}{b}{it}

\shorttitle{And yet it moves} 

\shortauthors{G\'omez et al.}

\begin{document}\title{And yet it moves: The
  dangers of artificially fixing the Milky Way center of mass in the presence of a massive Large Magellanic Cloud}

\author{Facundo A. G\'omez}\affil{Department of Physics and Astronomy,
  Michigan State University, East Lansing, MI 48824, USA}\affil{Institute
  for Cyber-Enabled Research, Michigan State University, East Lansing, MI
  48824, USA}\affil{Max-Planck-Institut f\"ur Astrophysik, Karl-Schwarzschild-Str. 1, D-85748, Garching, Germany}\email{fgomez@mpa-garching.mpg.de}
\author{Gurtina Besla}\affil{Steward Observatory, University of Arizona, 933 North Cherry Avenue, Tucson, AZ 85721, USA}  
\author{Daniel D. Carpintero}\affil{Facultad de Ciencias Astron\'omicas y Geof\'\i sicas, Universidad Nacional de La Plata, Argentina}\affil{Instituto de Astrof\'\i sica de La Plata, UNLP-Conicet La Plata, Argentina}  
\author{{\'A}lvaro Villalobos}\affil{Astronomical Observatory of Trieste, via G.B. Tiepolo 11, I-34143 Trieste, Italy}  
\author{Brian W. O'Shea}\affil{Department of Physics and Astronomy,
  Michigan State University, East Lansing, MI 48824,
  USA}\affil{Lyman Briggs College,
  Michigan State University, East Lansing, MI 48825, USA}
\affil{Institute for Cyber-Enabled Research, Michigan State
  University, East Lansing, MI 48824, USA}\affil{Joint
  Institute for Nuclear Astrophysics}
\author{and Eric F. Bell}\affil{Department of Astronomy, University of Michigan,
   830 Dennison Bldg., 500 Church St., Ann Arbor, MI 48109, USA}

\label{firstpage}

\begin{abstract}

Motivated by recent studies suggesting
that the Large Magellanic Cloud (LMC) could be significantly more massive than previously thought, we explore whether 
the approximation of an inertial Galactocentric reference frame is still valid in  the  presence  
of such a massive LMC. We find that previous estimates of the LMC's orbital period and apocentric distance 
derived assuming a fixed Milky Way are significantly shortened for models where the Milky Way is 
allowed to move freely in response to the gravitational pull of the LMC. Holding other parameters 
fixed, the fraction of models favoring first infall is reduced.  Due to 
 this interaction, the Milky Way center of mass within the inner 50 kpc can be significantly 
 displaced in phase-space in a very short period of time that ranges from 0.3 to 0.5 Gyr by as much as 30 kpc and 75 km/s.  
Furthermore, we show that the gravitational pull of the LMC and response of the Milky Way are likely to 
significantly affect the orbit and phase space distribution of tidal debris from the Sagittarius 
dwarf galaxy (Sgr). Such effects are larger than previous estimates based on  the torque of the LMC alone.
As a result, Sgr deposits  debris in regions of the sky that are not aligned with the present-day Sgr orbital plane.
In addition, we find that properly accounting for the movement of the Milky 
Way around its common center of mass with the LMC significantly modifies the angular distance between apocenters 
and tilts its orbital pole, alleviating tensions between previous models and observations. 
While these models are preliminary in nature, they highlight the central importance of accounting for the mutual gravitational interaction between the MW and LMC when modeling the kinematics of objects in the Milky Way and Local Group.

\end{abstract}


\section{Introduction}
\label{sec:intro}

Recently, a series of studies based on photometric, kinematic and dynamical arguments 
have enhanced our current understanding of the orbital history and mass of the 
Magellanic Clouds system \citep[see, e.g.][and references therein]{K13}. 
The results presented in these studies suggest 
that the Large Magellanic Cloud could be significantly more massive than previously thought. 
\citet{2010ApJ...721L..97B, 2012MNRAS.421.2109B} showed that 
the  observed  irregular morphology  and  internal  kinematics of  the
Magellanic  System  (in both the gas  and  stellar components)  are  naturally explained  by
interactions between  the Large and  the Small Magellanic  Clouds (LMC and SMC hereafter),  
rather  than  gravitational interactions  with  the  Milky Way (MW hereafter). 
\citet[][hereafter K13]{K13}  showed that in order for the SMC to be
bound to the LMC  for periods as large as 2 Gyr (the estimated age of
the  Magellanic stream) a LMC  with a  mass greater  than  $1 \times
10^{11}~M_{\odot}$ is required.  In addition, based on  proper
motion   measurements   obtained   using the  Hubble   Space   Telescope,
\citet{2007ApJ...668..949B} and K13 showed that for such massive LMC models, 
the Magellanic Clouds are likely to be experiencing their first infall towards the MW

Could  the acceleration of the inner regions of  the MW induced by 
such a massive LMC   be  significant, even if  it is experiencing
its first pericenter passage?  In binary stellar systems, the two stars orbit about a 
common center of mass that is often exterior to the more massive star.  The MW+LMC 
system may be analogous, where the center of mass of the combined system may be at a 
non-negligible distance from the Galactic center. A simple back-of-the-envelope calculation 
suggests that this may indeed be the case. For example, assuming  a MW model with a
dark matter halo of viral mass $M_{\rm vir} = 1 \times 10^{12}~M_{\odot}$, the
mass of the  MW enclosed within the LMC present-day position,  
$R_{\rm LMC} \approx 50$ kpc, is approximately $M_{\rm MW}^{50} \approx 3-4 \times 10^{11}~M_{\odot}$.
The LMC canonical model adopted by K13, based on the requirement that the 
LMC and SMC have been a long-lived binary, assumes a 
total mass of $M_{\rm LMC} = 1.8 \times  10^{11}~M_{\odot}$. In this MW + LMC system, the orbital
barycenter could be displaced  by as much as $\approx 14$ kpc from the  
Galactic center. The associated phase-space displacement of the MW with respect to its orbital
barycenter could  have a substantial impact on the inferred orbital properties 
of satellite galaxies, including the LMC itself. In other words, such a massive satellite orbiting the
MW at the present day  could pose a serious challenge to the commonly-adopted assumption of an inertial
Galactocentric reference frame.

While understanding the motion of the Milky Way and its neighbors is
of relevance for many Local Group studies, a deep understanding of the
expected response of the MW to the gravitational pull of such a
massive LMC is urgently needed for analyses based on orbital
integration using present-day phase-space coordinates as initial
conditions. Furthermore, due to  the extended  nature of  
the MW  stellar halo,  not all stars will experience the same acceleration 
from the LMC. This differential acceleration could introduce observable signatures on 
the phase-space distribution of extended tidal  streams, such as those associated with 
the Sagittarius dwarf galaxy (Sgr). The Sgr tidal tails span at least $300^{\circ}$  across the
sky   \citep{1997AJ....113..634I},   and    have   been   observed   at
Galactocentric     distances     as     large     as     $100$     kpc 
\citep[e.g][]{2003ApJ...599.1082M,2003ApJ...596L.191N,2011ApJ...731..119R,2013ApJ...765..154D}. 
Indeed,  \citet{2013ApJ...773L...4V} showed that the torque on 
Sgr exerted by the LMC can introduce non-negligible  perturbations to the orbit of
Sgr  and its  distribution of  debris.  Their work, however, considered an 
spatially-fixed MW model, thus neglecting the dynamical response of the MW 
to the gravitational pull of the LMC. 

The aforementioned perturbations, associated with the plausible presence of a massive LMC, could 
even  influence the determination  of the  present-day Galactic mass  distribution.
Multiple observational programs have provided, and will continue to provide, very accurate   photometric,   
astrometric, and spectroscopic information for enormous samples of stars, not only 
in the Galactic disk but also in the more extended stellar halo 
\citep[see e.g.][]{sdss,gaia,rave,2007PASA...24....1K,segue,hermes,lamost,2012Msngr.147...25G}. 
During the last two decades several  studies were devoted  to the development and application of powerful  theoretical 
and statistical tools that could allow  us  to efficiently  mine  these observational data sets. An important goal in 
many of these studies is to statistically infer the present-day Galactic mass  distribution. 
It is customary for these studies to consider as input data dynamically  young  and extended  stellar streams 
\citep[e.g.][]{2004ApJ...610L..97H,2005ApJ...619..800J,2010ApJ...714..229L,2010ApJ...712..260K,
2012MNRAS.424L..16L,2013MNRAS.433.1826S,2013ApJ...773L...4V,2014ApJ...795...94B,2014MNRAS.439.2678D,
2014MNRAS.445.3788G,2014ApJ...794....4P}. The  reason behind  
this choice is  simple: these types of spatially extended streams are  expected to  approximately  
delineate the  orbit of  their corresponding progenitors in phase-space 
\citep[see][]{2011MNRAS.413.1852E,2013MNRAS.433.1813S}. 

For simplicity, in most of these works the
MW's mass distribution has been assumed to be smooth and static, not only  structurally   but  also  spatially.  
Assuming a  frozen-mass potential may not strongly affect the results of 
these analyses.  The MW's mass is not  expected to have significantly evolved during 
 the last 2 to 3 Gyr \citep[e.g.][]{2005ApJ...635..931B}, a dynamical timescale that  pertains to these 
studies.  On the other hand,  the assumption that the MW can be regarded as an inertial  
frame has not been thoroughly tested. If  one neglects  the presence  of the  
LMC, the  MW's accretion
activity can be regarded as  quiescent during this period of time.  
However, the degree to which the presence of a massive LMC could significantly 
affect the statistically-inferred parameters that best describe the Galactic potential remains to be studied.

The dangers associated with artificially fixing the MW center of mass 
have been considered by several authors in the past. 
One of the first works to explore this was presented  
by \citet{1983ApJ...274...53W}.  Using $N$-body
simulations, this study showed  that  the  orbital  decay   rate  of  a  satellite  galaxy  is
artificially  enhanced by  fixing the  host center  of  mass.  Current
analytic prescriptions  to model dynamical friction are  fine-tuned by
calibrating against results of fully self-consistent $N$-body
simulations  \citep[e.g.,][]{1997MNRAS.289..253C,  2004MNRAS.351.1215B,
  2005A&A...431..861J}. More  recently,  \citet{2014ApJ...789..166P}
 discussed the effects that  the time evolution of the orientation of the disk angular
 momentum vector  with respect to an initial reference frame
could  have  on  {\it  Gaia} measurements.   Such perturbations to 
the disk angular momentum could be caused by, e.g., the    time-dependent
accretion of  gas \citep{2006MNRAS.370....2S, 2010MNRAS.408..783R},
the   predicted   tumbling   of   the  Galactic   dark   matter   halo
\citep[e.g.][]{2005ApJ...627..647B,                2007MNRAS.380..657B,
  2011MNRAS.416.1377V} and by the tidal interaction of a fairly massive  
LMC \citep{2012MNRAS.422.1957B}.

In this work we revisit the problem of a non-inertial MW reference frame 
by modeling the interaction between the MW, the Sagittarius dwarf galaxy (Sgr hereafter), 
and a LMC that is undergoing its first infall at the present day.
We focus our analysis on two possible situations
where  the response of the  MW to the  gravitational pull of
the  LMC could  induce significant  perturbations: namely, the inferred orbit of the LMC about the MW, and the orbit and  tidal debris from the Sgr dwarf galaxy.
To this end, we use a variety of different techniques to model the gravitational interaction between these
three galaxies. In Section~\ref{sec:lmc_mass} we provide a justification for the LMC mass range explored in our experiments.
In Section~\ref{sec:lmc} and~\ref{sec:sgr} we use smooth analytic representations of the Galactic
potentials to characterize the significance of this perturbative effect on the orbital properties of both the 
LMC and Sgr. In Section~\ref{sec:nbody_sgr} we use full $N$-body simulations to explore the consequences of 
a non-inertial Galactocentric reference frame on the phase-space distribution of the Sgr tidal debris.  We  conclude   
and  discuss  our  results  in Section~\ref{sec:conc}.

\section{The Mass of the LMC}
\label{sec:lmc_mass}

In this study we consider LMC mass models that range from 
$3 \times 10^{10}$ to $2.5 \times 10^{11}~M_\odot$. We would like to stress that the considered LMC masses are meant to
represent the total LMC infall mass up to its virial radius, as opposed to the present-day observational constraint 
within its optical radius. In this section 
we justify this mass range and explain why high-mass LMC models are currently
favored. We refer the reader to \citet{Bproc} for a more extended discussion.

Our goal is to  explore  the effects of a massive LMC on
the  assumption that the  MW can  be considered  an inertial  frame of
reference. The  mass of  the LMC is  the dominant uncertainty  in the
orbital   history  of   the Magellanic Clouds since  dynamical   friction  is
proportional to its mass. Moreover, the  mass of  the LMC  also 
controls the  orbit of  the Small Magellanic Cloud,
ultimately determining how long  the two galaxies have interacted with
each other as a binary pair.

Observationally, the total mass of the LMC is only constrained within 
the optical radius.  The LMC has a well-defined rotation curve that peaks at 
$V_{\rm circ} = 91.7 \pm$ 18.8 km/s and remains flat out to at least 8.7 kpc 
\citep{2014ApJ...781..121V}. 
This peak velocity places the LMC squarely on the well-defined
baryonic Tully-Fisher relation \citep{McG12}. 
This implies a minimum enclosed total mass of 
$M$(8.7 kpc) $= 1.7  \times 10^{10}~M_\odot$, and further implies that
the LMC is dark matter-dominated. The total mass of the LMC may be much 
larger than this, depending on the tidal radius.  

There is strong evidence that the stellar disk of the LMC extends to
15 kpc \citep{2009IAUS..256...51M,Saha10}. If the rotation curve stays flat to at least this 
distance then the total mass enclosed is $M$(15 kpc) =
 $V_{\rm circ}^2r/G  \sim 3 \times 10^{10}~M_\odot$. 
This minimum value is consistent with LMC masses adopted by 
traditional models of the orbital evolution of the Magellanic Clouds \citep[e.g.][]{MF80,GN96}. 
Note however that this estimate only takes into account the total mass of the LMC within 15 kpc. 
Thus, it may significantly underestimate its total infall mass within its virial radius; this is the 
quantity of interest for this work.

The total dynamical mass of the 
LMC at infall, up to its virial radius, can be estimated using its baryon fraction. Currently, the LMC has 
a stellar mass of $2.7\times10^{9}$ $M_\odot$ and a gas mass 
of $5.0 \times 10^{8}$ $M_\odot$. The baryonic mass 
of the LMC is thus $M_{\rm bar} = 3.2 \times 10^{9}~M_\odot$. 
Using the minimum total mass of  $M_{\rm tot} = 3\times10^{10}~M_\odot$, 
the baryon fraction of the LMC becomes $M_{\rm bar}$/$M_{\rm tot}$ = 11\%. 
This is much higher than the baryon 
fraction of disks in galaxies like the MW, which is of the order of 
3-5\%. In the shallower halo 
potentials of dwarf galaxies, stellar winds should be more 
efficient, making baryon fractions even lower, not higher.  

This analysis is further complicated if material has been removed from the
LMC.  \citet{Fox14} have recently estimated the total 
gas mass (HI and ionized gas) outside the Magellanic Clouds at
$2\times10^{9} (d/55 {\rm kpc})^2~M_\odot$, with $d$ the distance to the Magellanic stream. 
If half of this material came from the LMC, as suggested by \citet{Nidever08},
its initial baryon fraction would be 14\% -- approaching the cosmic value.
Note that the bulk of the Magellanic Stream likely resides at distances 
of order $d=$100 kpc, rather than 55 kpc, in which case the baryon fraction would 
increase to $\sim$20\%. 

In order for the baryon fraction to match observational expectations of 
$f_{\rm bar} \sim$3-5\%, the total mass of the LMC (at least at infall) 
needs to have been $6-20 \times 10^{10}~M_\odot$. 
This higher total mass is consistent with cosmological expectations from 
halo occupation models that relate a galaxy's observed stellar mass 
to its halo mass.  Using relations from \citet{Mos13}, the mean halo mass for a 
galaxy with a stellar mass of $2.7 \times 10^{9}~M_\odot$ is 
$1.7 \times 10^{11}~M_\odot$, implying a baryon fraction of $f_{\rm bar}\sim$ 2-4\%.
 Because there is a large scatter in halo occupation models, we have considered a maximal halo 
mass for the LMC of $2.5 \times 10^{11}~M_\odot$.

The halo occupation  model relations are primarily invoked to motivate initial
conditions for a first infall model. As shown in K13 and later in this work, first infall models are obtained in 
 Milky Way-like hosts with a total mass $\approx 1\times10^{12}$ M$_{\odot}$, regardless of the total LMC mass (within the range considered here). If the Clouds have only recently been accreted there has not been enough
time to severly truncate the LMC halo and, as a result, its current mass should approximately reflect its
infall mass; i.e., the mass the LMC halo had upon first crossing the virial radius.
Note as well that high-mass LMC models, $> 10^{11}$ M$_{\odot}$, are necessary 
in models of the formation of the Magellanic Stream as they allow for a long-lived ($\sim$ 4 Gyr) 
LMC/SMC binary configuration.  The relative velocity between the Clouds is $\sim$ 130 km/s;  
high-mass LMC models (masses of order $10^{11}$ M$_\odot$)  are needed 
to explain how the LMC can have held on to the SMC if it is moving at such speeds. This argument 
has been outlined in K13. Based on their parameter space search, and the requirement
that the LMC and SMC have been a long-lived binary, we adopt a canonical mass model for the LMC $= 1.8 \times 10^{11}$ M$_{\odot}$.

A very important uncertainty in the kind of analytic orbital integration
schemes we employ in this study is the mass evolution of the LMC over time. 
The arguments laid out in this section are for the required {\it infall}
mass of the LMC. The subsequent mass loss incurred by the LMC 
as it orbits about the MW will necessarily cause
significant modifications in the orbits presented in the following section.  
A proper analysis accounting for this effect requires detailed
$N$-body simulations that, in principle, are beyond the scope of the simple backward integration 
scheme presented here. Nonetheless, to validate our assumptions, we will compare the results from one of the
MW+LMC mass combinations with those obtained with the corresponding fully self-consistent $N$-body model.

\section{The orbit of the LMC about the MW}
\label{sec:lmc}

Our goal in this Section is to explore whether artificially fixing the
MW center  of  mass could have significant implications on the  
inferred orbital  properties of  the  LMC. For  this purpose,  we will
integrate  the orbits of different LMC models backwards in time in  MW-like hosts
that are kept artificially fixed in space and that are also allowed to react to
the gravitational pull of the LMC. In order to make a direct comparison with the results presented
in K13 we will start by considering  smooth, analytic representations of the Galactic potentials. We will later compare our results
with those obtained from full $N$-body simulations. The models and methodology 
are described in Section~\ref{sec:analy_lmc_mod}. Our results are presented on Section~\ref{sec:mw_lmc_ana}.

\subsection{Analytic models}
\label{sec:analy_lmc_mod}

\begin{figure}
\centering
\includegraphics[width=80mm,clip]{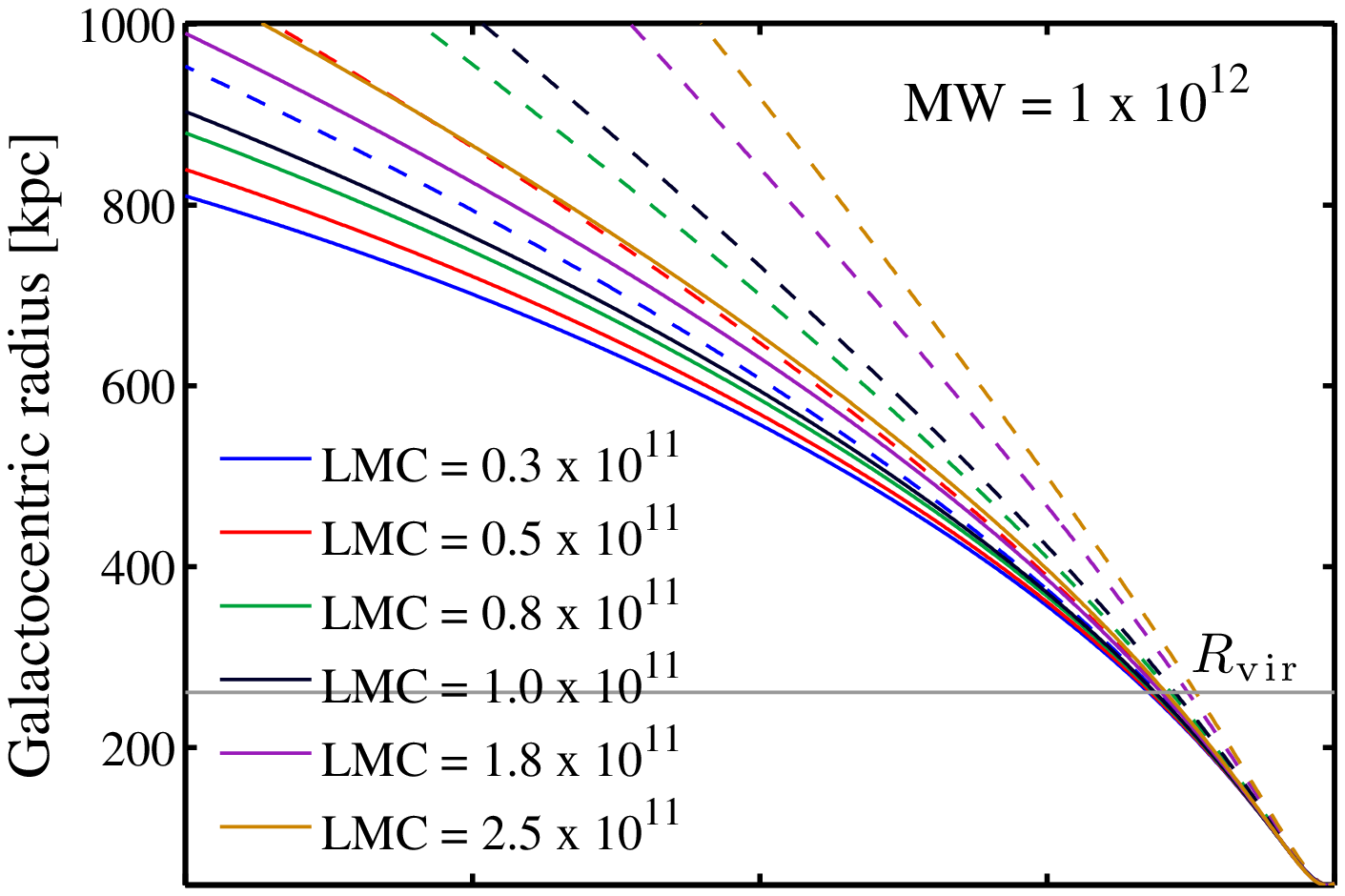} \\
\includegraphics[width=80mm,clip]{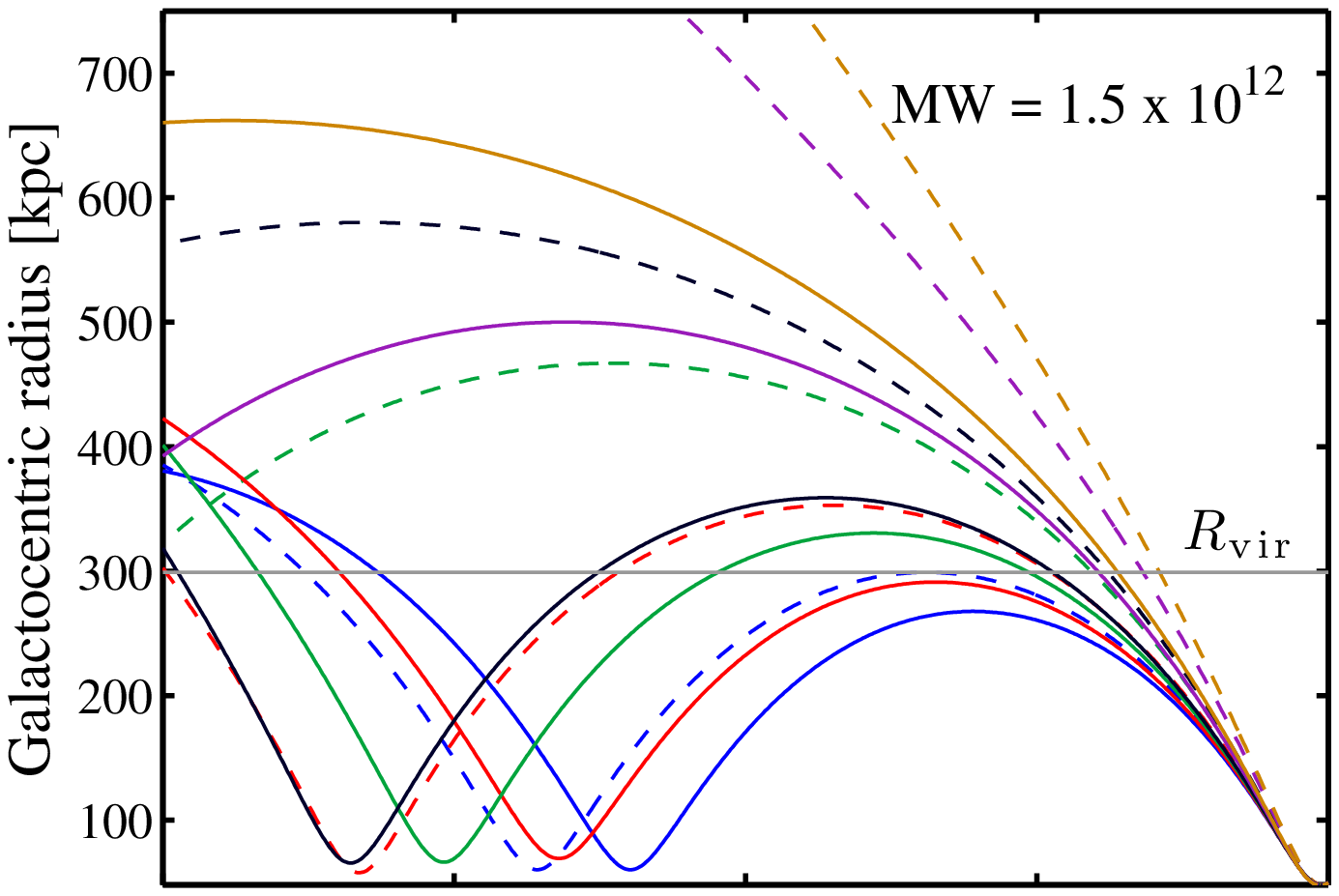} \\
\includegraphics[width=80mm,clip]{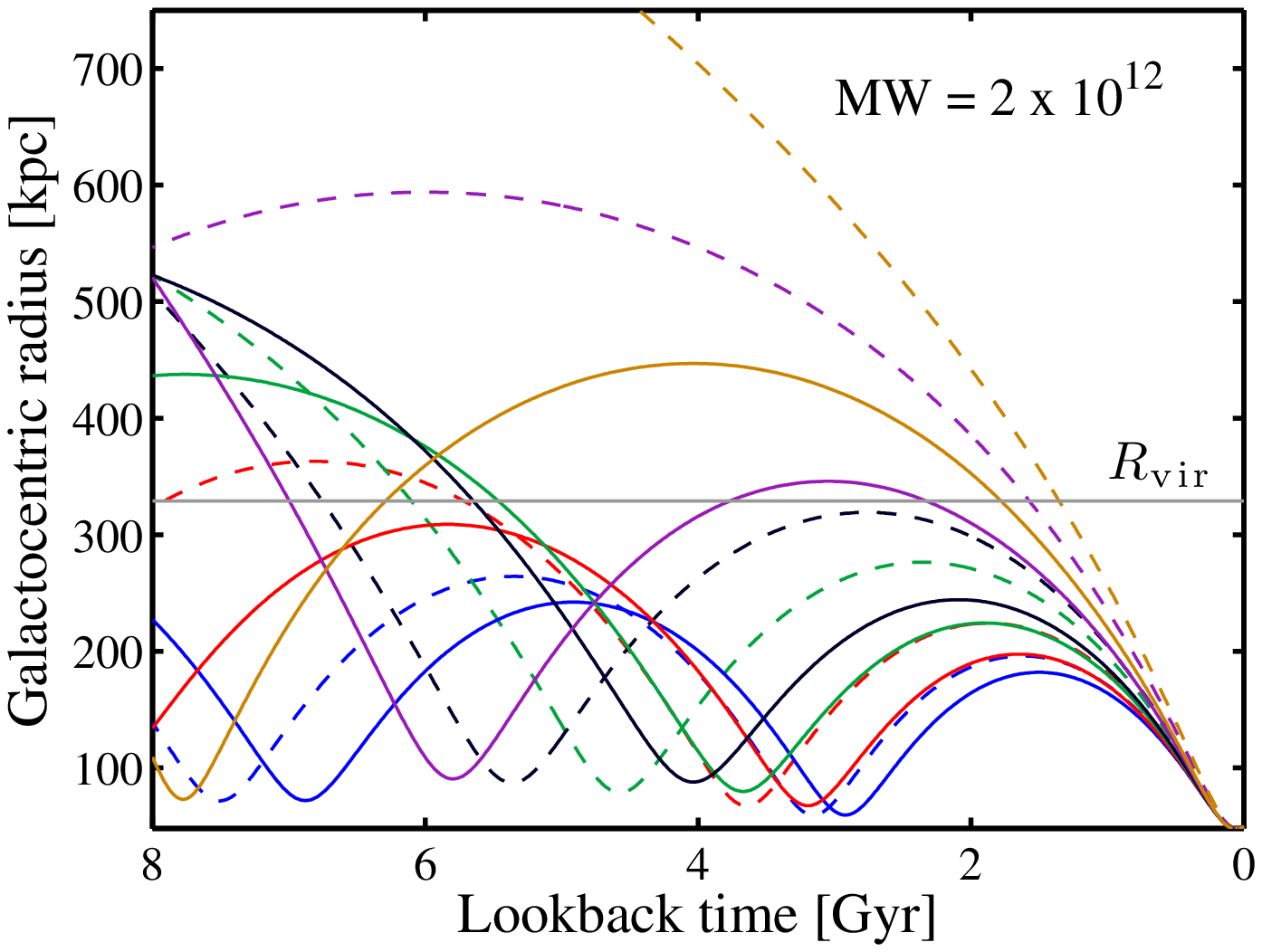}
\caption{Time evolution of the  galactocentric radius of different LMC
  models  in  three  MW-like  host  potentials.   Orbits  are
  integrated backwards  in time.  Note  that $t=0$ Gyr  corresponds to
  present-day.   The different  color-coded  lines  show the  results
  obtained with different  LMC models, as indicated in  the top panel.
  From top to bottom, the MW models have $M_{\rm vir}
  = 1,~1.5$ and  $2\times 10^{12}~M_{\odot}$, respectively. The dashed
  lines show  the results  obtained in MW models in  which the
  center of  mass has been  artificially fixed.  Solid lines  show the
  results  obtained  when  the  host   is  allowed  to  react  to  the
  gravitational pull exerted by the  LMC. Note the shorter LMC orbital
  periods obtained in the latter case.  The more massive the LMC, the
  larger  the  change  in  orbital  period.}
\label{fig:lmcs}
\end{figure}

\subsubsection{Methodology}
\label{sec:lmc_meto}

To follow  the evolution of the gravitational  interaction between the
MW and  the LMC  we  used a  symplectic
leapfrog integration scheme \citep{2001NewA....6...79S}. Both the host
and  the  satellite are  represented  with  analytic potentials;  the
center  of  each  one  follows  the orbit  that  results  from  the
acceleration of the  other. In practice, this is done by assigning to the center of mass of each 
galaxy a mass-less tracer particle. The orbit of each tracer particle is determined by the smooth gravitational potential 
associated with the secondary galaxy. If, as in Section~\ref{sec:sgr}, a third galactic model is included, 
the orbit of each tracer particle will be determined by the smooth and non-trivial potential associated with the overlapping 
density distributions of the two remaining galactic models. Note that, even though we use mass-less particles 
as phase-space tracers of the galactic centers of mass, we assign to each galactic model a 
spatially extended density distribution (see Section~\ref{sec:pot_gal_smooth}). 
Thus, at any given time, the acceleration exerted by the LMC on the 
MW (and vice versa) is computed by only taking into account the mass that is enclosed within a sphere centered 
on the LMC (and vice versa) of radius equal to the distance between the two center of masses. In all 
cases the  orbits are integrated backwards from their present-day positions and velocities.

As in \citet{2007ApJ...668..949B} and K13, we ignore the mass evolution  of the LMC
owing to the MW's tidal field. In addition, we do not
follow the time evolution of the mass or the structural parameters of
the MW potential. Since the potentials considered are structurally
frozen,  there is  no dynamical  friction exerted  on the satellite galaxies. 
Therefore, we  model  this acceleration  using  an
approximation  of Chandrasekhar's  dynamical  friction  formula
\citep{C43,BT08},
\begin{equation}
\label{eq:chandra}
\frac{{\rm d}{\bf v}_2}{{\rm d}t} =
-4 \pi G^2 M{_2} \rho_1 \ln\Lambda \left[ \int_{0}^{v_{2}} v^2
    f_{1}(v){\rm d}v \right]\frac{{\bf v}_2}{v_2^3},
\end{equation}
where the  subindex 1 refers to  the galaxy causing  the friction, the
subindex  2  to the  galaxy  being  decelerated,  {\bf v}$_2$  is  the
relative  velocity  of both  interacting  galaxies, $M$ is the mass of the corresponding galaxy, 
$\rho$ the mass density,  $f$  the distribution function of velocities, $G$ the gravitational 
constant and  $\Lambda$ is  the Coulomb  factor. For simplicity, in these experiments we neglect the 
dynamical friction exerted by the LMC on the MW. 

Under  the assumption  of  a Maxwellian
velocity    distribution and a constant background density field, it is possible to   
approximate   the    integral    in Equation~\ref{eq:chandra} by:
\begin{equation}
\label{eq:chandra_ap1}
\int_{0}^{v_{2}} v^2
f(v){\rm d}v \approx \left( {\rm erf}(x)- \frac{2x}{\sqrt{\pi}} {\rm e}^{-x^2} \right),
\end{equation}
where $x  = v_{2}/\sqrt{2}\sigma$ \citep{BT08}.  Here, 
$\sigma$  is the one-dimensional velocity  dispersion  of the  host  dark  matter  halo.  
We  adopt  an analytic approximation  of $\sigma$ for  an NFW profile as  derived by
\citet{2003ApJ...598...49Z}.  Following \citet{2007ApJ...668..949B} and K13, for these  experiments 
we consider a value of the Coulomb factor that varies as a function of the 
satellite's galactocentric distance as described by \citet{2003ApJ...582..196H}.
The \citet{2003ApJ...582..196H}  Coulomb factor not only scales as a function of the satellite's 
separation to the host but also as a function of the satellite's scale radius.  
As in K13, a fixed scale radius of 3 kpc is assumed in all cases. This may possibly overestimate 
the role of dynamical friction in the orbital history of high mass LMC models. Detailed comparisons with 
$N$-body models are required to properly estimate the degree of error, which will be 
complicated by mass loss owing to MW tides and the presence of the SMC. Such 
an analysis is beyond the scope of this study. Nonetheless, as we will later show in this section, such effects 
are very small when considering LMC models that are currently undergoing their first pericenter passage. 
Moreover, our goal is to assess the 
effects of a non-inertial frame of reference on the LMC's orbit rather than to determine 
the exact orbital history itself. As such, this methodology will sufficiently illustrate 
the general change in the orbits.   If indeed dynamical friction is overestimated for 
high mass LMC models, their orbits will be less eccentric backwards in time, augmenting 
the perturbative effects we illustrate here.

\subsubsection{Galactic potentials}
\label{sec:pot_gal_smooth}

To model  the MW  potential we choose a  three-component system,
including a Miyamoto-Nagai disk \citep{1975PASJ...27..533M}
\begin{equation}
\label{eq:mn}
\Phi_{\rm disk}=-\frac{GM_{\rm disk}}{\sqrt{R^{2}+\left(r_{\rm
      a}+\sqrt{Z^{2}+r_{\rm b}^{2}}\right)^{2}}},
\end{equation}
a Hernquist bulge \citep{H90},
 \begin{equation}
\label{eq:hernq}
\Phi_{\rm bulge}=-\frac{GM_{\rm bulge}}{r+r_{\rm c}},
\end{equation} 
and a  Navarro, Frenk  \&  White  dark matter  halo
\citep[][hereafter, NFW]{1996ApJ...462..563N}
\begin{equation}
\Phi_{\rm halo}=-\frac{GM_{\rm
    vir}}{r\left[\log(1+c)-c/(1+c)\right]}\log\left(1+\frac{r}{r_{\rm s}}\right).
\end{equation}

\begin{deluxetable}{llllllll}
  \tabletypesize{\footnotesize} \tablecaption{Parameters  of the MW-like  
  potential  used  in  our  simulations.   \label{tab:mw}}
  \tablewidth{230pt}      \tablehead{\colhead{$M_{\rm     vir}$}     &
    \colhead{$R_{\rm vir}$} & \colhead{$r_{\rm s}$} & \colhead{$M_{\rm
        disk}$}  &  \colhead{$r_{\rm a}$}  &  \colhead{$r_{\rm b}$}  &
    \colhead{$M_{\rm bulge}$} & \colhead{$c_{\rm bulge}$}} \startdata
  $100$ & $261$ & $26.47$ & $6.5$ & $3.5$ & $0.53$ & $1.0$ & $0.7$  \\
  $150$ & $299$ & $31.27$ & $5.5$ & $3.5$ & $0.53$ & $1.0$ & $0.7$ \\
  $200$ & $329$ & $35.15$ & $5.0$ & $3.5$ & $0.53$ & $1.0$ & $0.7$
\enddata
\tablecomments{Masses  are  in  $10^{10}~M_{\odot}$ and  distances  in
  kpc. The  scale radius for  the  Hernquist profile dark matter halos 
  are obtained from $r_{\rm s}$ through equation~\ref{eq:nfwtohern}}
\end{deluxetable}

\begin{deluxetable}{lllllll}
  \tabletypesize{\footnotesize}  \tablecaption{Parameters  of the  LMC
    models    used   in    our    simulations.    \label{tab:lmc}}
  \tablewidth{200pt}      \startdata 
  \hline \hline \vspace{-0.1cm} \\
  $M_{\rm LMC}~[10^{10}~M_{\odot}]$  & 3 & 5 & 8 & 10  & 18 & 25 \\
  $r_{\rm LMC}$ [kpc] & 8 & 11 & 14 & 15 & 20 & 22.5  
\enddata
\tablecomments{These parameters are used for both Plummer and 
  Hernquist profiles.}
\end{deluxetable} 

Here, $R$ and $Z$ are the radial and vertical cylindrical coordinates and $r$ is the radial 
spherical coordinate. The dark matter (DM) halo viral mass, $M_{\rm vir}$, is defined as the mass enclosed 
within the radius where the dark matter density is 360 times the average matter density
\citep{2012ApJ...753....8V}. In  all models  the  disk scale  length  and height,  $r_{\rm a}$  and
$r_{\rm b}$,  are kept fixed at  3.5 and 0.53  kpc, respectively.  The
bulge mass and scale radius, $M_{\rm bulge}$ and $r_{\rm c}$, are also
kept fixed at $10^{10}~M_{\odot}$ and 0.7 kpc, respectively. In addition, the NFW density profiles are 
truncated at the virial radius.  The  remaining parameters  are allowed  to vary  in order  to explore
different  models for  the  MW potential. The adiabatic contraction of 
the dark matter halo associated with the presence of a disk was taken into account 
using the CONTRA code \citep{2004ApJ...616...16G}. The  values of  the parameters    
for the different models are listed in Table~\ref{tab:mw}. The circular velocity curve 
of these Galactic models is shown in Figure 8 of K13. Note that in  all cases the circular 
velocity at the  solar  circle, $R_{\odot}  \approx  8.3$  kpc, takes a  value  of
$V_{\odot} = 239$ km/s \citep{2011MNRAS.414.2446M}. Note that a lower value of $V_{\odot}$, 
e.g. $V_{\odot} = 218 \pm 6$ km/s \citep{2012ApJ...759..131B}, implies a lower mass MW model within 
its inner regions. Thus, for such Galactic models, any plausible two-body interaction with our LMC 
models would be enhanced.

The LMC is modeled using a Plummer
sphere \citep{1911MNRAS..71..460P},

 \begin{equation}
\Phi_{\rm LMC}=-\frac{GM_{\rm LMC}}{\sqrt{r^{2}+r_{\rm LMC}^{2}}}.
\end{equation}

Following K13, a variety  of LMC masses are  explored, ranging from $M_{\rm  LMC} = 3
\times  10^{10}$ to $2.5  \times 10^{11}  M_{\odot}$. A detailed justification
for this explored LMC mass range was provided in Section~\ref{sec:lmc_mass}. 
The parameters  that describe
these  LMC models are  listed in  Table~\ref{tab:lmc}.  Note  that the
scale  radius  of  each model  is  chosen  such  that the  total  mass
contained within 9 kpc is $\approx 1.3 \times 10^{10}$ M$_{\odot}$, as
indicated by the LMC's rotation curve \citep{2009IAUS..256...81V, 2014ApJ...781..121V}.

\subsection{An interacting MW + LMC model}
\label{sec:mw_lmc_ana}

\subsubsection{The orbital properites of the LMC}
  
To explore whether artificially fixing the MW's center of mass could  
have a significant effect on the inferred orbital properties of the LMC, 
we generate LMC-like orbits by integrating the galaxies backwards in time from their present 
day phase-space coordinates.
The initial orbital conditions for  all LMC models, in a Galactocentric
reference  frame, are $(X,Y,Z)_{\rm  LMC} =  (  -1, ~-41,  ~-28)$ kpc  and
$(v_{x}, ~v_{y}, ~v_{z})_{\rm LMC} = (-57,~-226,~221)$ km/s. The quoted velocities represent the mean 
 value of LMC's velocity and are obtained from K13.   Velocities are based on proper  motion measurements
obtained with three  epochs of Hubble Space Telescope  data.

The color-coded dashed lines  in Figure~\ref{fig:lmcs} show  the time
evolution  of  the LMC's  galactocentric  distance  in MW-like
models where the center  of mass has been artificially nailed down.
From  top  to  bottom,  the  different panels  show  the  results  of
integrating these  orbits over  a period of  8 Gyr in  Galactic models
with dark matter halo masses of  $M_{\rm vir} = 1,~1.5,$ and $2 \times
10^{12}~M_{\odot}$,  respectively.   As   expected,  our  results  are in
very good agreement  with those found by  K13 (see Fig. 11 of K13). 

In a MW model with $M_{\rm vir}  = 1 \times 10^{12}~M_{\odot}$
(top panel of Figure~\ref{fig:lmcs}), the  resulting LMC-like orbits show periods  
larger than a Hubble time, $T_{0} \approx  13.73$ Gyr, independent of the satellite's
total mass.   Increasing the mass of  our MW  models results in
shorter  LMC orbital  periods.  As  a result,  the less  massive LMC
models start  to show more than  one pericenter passage  within 8 Gyr.
Note  that, as  discussed by  K13, orbits  with periods,  $P$, between
$T_{0} > P \gtrsim T_{0}/2$ Gyr may not be physical since, according to a
general timing argument, 
the LMC  must have had a pericentric approach
with the MW very early on, at  the time of the Big Bang \citep{1959ApJ...130..705K,2008MNRAS.384.1459L}.  
Thus, if there 
has been more than one complete orbit, the LMC period must  be  
$< T_{0}/2  \approx  6.9$ Gyr. Although this argument suffers from  
oversimplification, it provides a rough estimate of the largest possible 
LMC orbital period.  In a  MW-like
host with a mass $M_{\rm vir} = 1.5 \times 10^{12}~M_{\odot}$
(middle  panel),  LMC models  with  masses  $M_{\rm  LMC} \leq  5  \times
10^{10}~M_{\odot}$ have completed a  full orbital period within 6.9 Gyr.
For  our  most  massive  MW  model ($M_{\rm  vir}  =  2  \times
10^{12}~M_{\odot}$,  bottom panel)  only  the two most  massive LMC  models
($M_{\rm LMC} = 2.5$ and $1.8 \times 10^{11}~M_{\odot}$) exhibit orbital periods
longer than 6.9 Gyr, in agreement with K13. 

\begin{figure}
\centering
\includegraphics[width=80mm,clip]{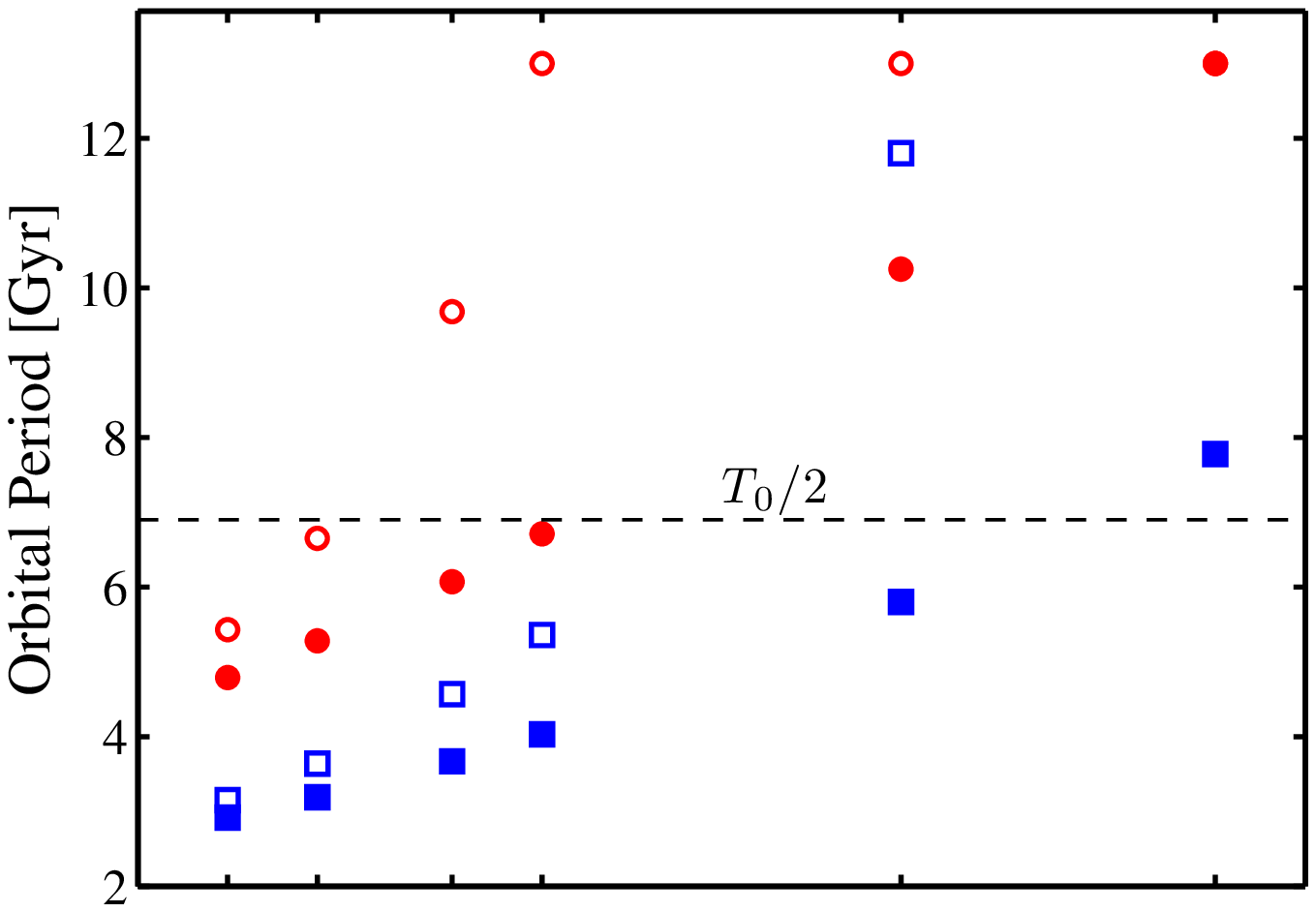}\\
\includegraphics[width=80mm,clip]{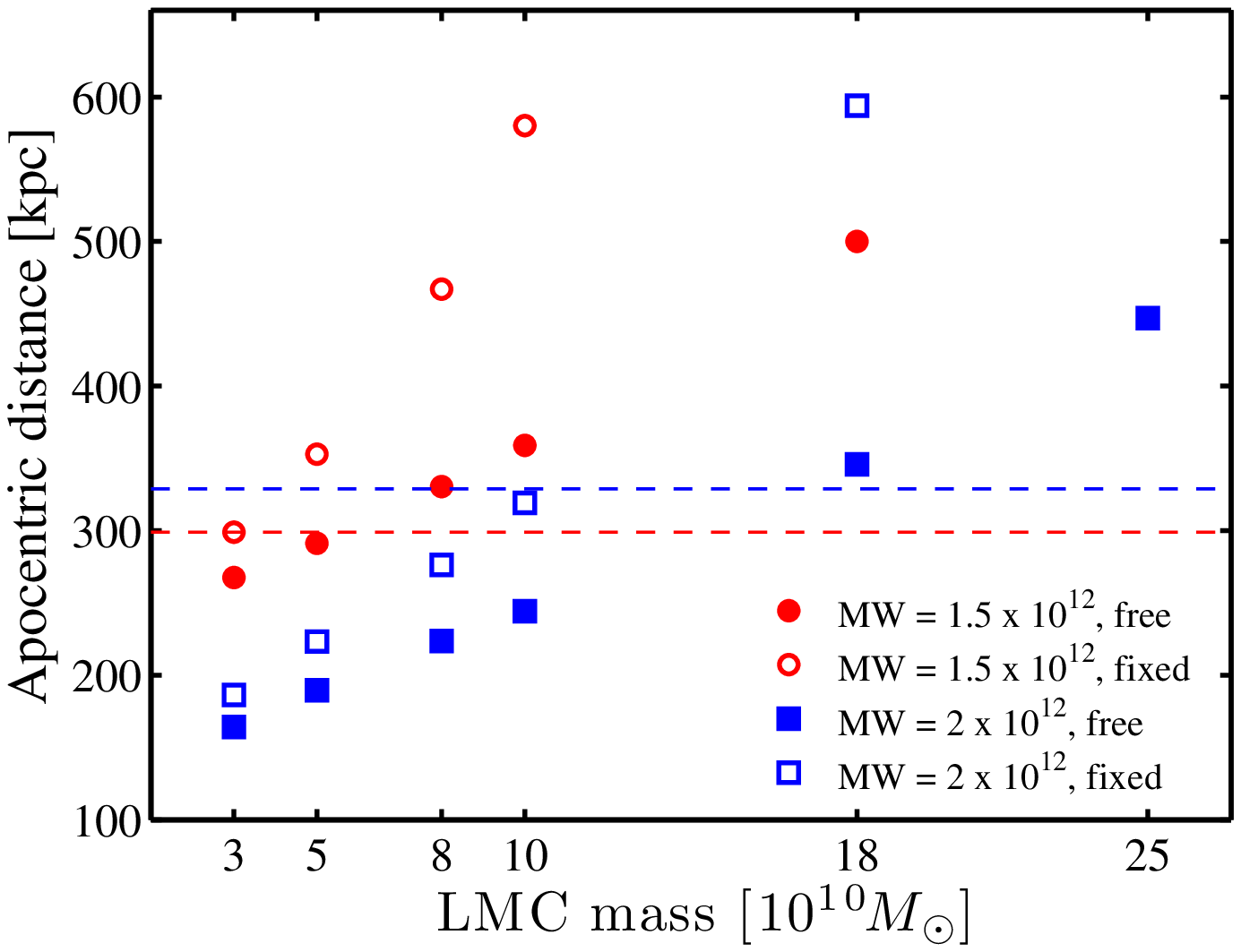}
\caption{Top panel: Orbital period of  LMC-like orbits as a function of the LMC total mass. 
The red and blue symbols indicate the period of orbits integrated in MW-like host 
with $M_{\rm vir}  = 1.5$ and $ 2 \times 10^{12}~M_{\odot}$, respectively. Open symbols show the orbital periods
obtained when the MW is artificially fixed in space. 
Filled symbols show the results obtained when the MW is allowed to freely
react to the gravitational pull exerted by the LMC. The black dashed line indicates half of the Hubble time,
$T_{0}/2$. Bottom panel: As in the top panel, for the apocentric distance of the different LMC-like orbits.
The red and blue dashed lines indicate the virial radius of the MW-like host with 
$M_{\rm vir}  = 1.5$ and $ 2 \times 10^{12}~M_{\odot}$, respectively. The orbital properties shown in this figure 
were obtained using the mean LMC's velocity presented in K13.}
\label{fig:period_apo}
\end{figure}

The color-coded solid lines in Figure~\ref{fig:lmcs}  show the orbits
of the same  LMC models, now in MW potentials that are allowed
to  freely react to  the gravitational  pull of  the LMC.   It becomes
abundantly  clear  that  nailing down  the  MW  center  of mass  has  a
very significant effect on the  backward time integrated orbits, particularly for
the most  massive LMC models. In all cases, the  orbital periods and
apocentric distances are significantly  shorter. As the mass of the LMC
becomes larger, and thus more comparable to the MW mass enclosed within the LMC's location,
the two-body interaction becomes more relevant. In other words, the more massive the 
LMC model, the more significant the changes in the resulting orbits are.
This can  be inferred from the orbits shown  in, e.g., the middle  panel of  Figure~\ref{fig:lmcs}
($M_{\rm vir} =  1 \times 10^{12}~M_{\odot}$). The acceleration experienced by the MW towards the LMC, and
the corresponding displacement  of its  center of  mass, result  in both  a  shorter LMC
orbital period and a smaller apocentric distance in a Galactocentric reference frame. 

Note that  this change in orbital period  is not related to
the artificial enhancement of dynamical friction discussed by \citet[][W83]{1983ApJ...274...53W}.
Using $N$-body simulations, W83 finds that artificially fixing the host's center of 
mass results in more efficient dynamical friction than when the host is allowed 
to orbit. The reason for this behavior is attributed to the different global patterns excited 
by the orbiting satellite on the density distribution of the host \citep[for a detailed discussion about this subject see][]{1997MNRAS.289..253C}. As opposed to the $N$-body models considered 
in W83, the galaxies in our  analysis are modeled through analytic and  
structurally frozen potentials.  Thus, the perturbation of 
the satellite cannot generate wakes in the host's density field. 
Changes shown in Figure~\ref{fig:lmcs} are  mainly a  reflection of the  
resulting orbits about the barycenter of the system. Note however that, due to 
the shortening of the satellites' orbital periods and apocentric distances, 
dynamical friction would act more efficiently in the free MW models than in the fixed MW models.

The orbits of the  LMC models in a MW-like host  with a mass of
$1.5  \times  10^{12}~M_{\odot}$ are  shown  in  the  middle panel  of
Figure~\ref{fig:lmcs}. Even in this  more massive host, the effects of
``freeing'' the MW are still very significant.  Now, all LMC
models with $M_{\rm LMC} \leq 1 \times 10^{11}~M_{\odot}$  have completed  a full  orbit within  $6.9$  Gyr.  In a   MW-like  host   with   a   total   mass  of   $2   \times
10^{12}~M_{\odot}$  (bottom panel)  all but the most massive LMC model have  completed a  full
orbit within 6.9  Gyr. 

We summarize and quantify the changes in our LMC-like orbits 
in Figure~\ref{fig:period_apo}. The top panel shows the orbital periods of the most recent
orbit, obtained in both free (filled symbols) and fixed (open symbols) MW-like models. 
We focus only on those orbits that have completed at least one orbit about the MW within a Hubble time.
This figure clearly shows how dramatic the change on the orbital period can be, especially for the most
massive LMC models. For example, for a MW-like host with $M_{\rm vir} = 1.5 \times 10^{12}~M_{\odot}$ and 
a LMC model with $M_{\rm LMC} = 1 \times 10^{11}~M_{\odot}$, the period changes from $13.1$ Gyr $\approx T_{0}$ 
(fixed MW) to 6.8 Gyr $\approx T_{0}/2$ (free MW). The bottom panel shows the corresponding changes in the 
apocentric distance, $R_{\rm apo}$, as a function of LMC mass. Note that for the  MW + LMC mass model combination 
discussed above, the apocenter goes from $R_{\rm apo} \approx 1.94 R_{\rm vir}$ to 
$R_{\rm apo} \approx 1.2 R_{\rm vir}$. The change in the inferred orbital properties of our 
LMC-like models suggests that, even though a first-infall 
is still a very plausible scenario, the limiting LMC-MW mass combinations 
that could host a first infalling LMC are noticeably affected; 
it raises the required minimum LMC mass and disfavors MW models with $M_{\rm vir} \geq 1.5 \times 10^{12}~M_{\odot}$.

\subsubsection{Phase-space displacement of the MW center of mass}

We have illustrated that  the presence of a massive LMC can substantially alter the orbital 
barycenter of the MW + LMC system even in a first infall scenario. It is thus interesting to quantitatively
characterize the displacement of the MW center of mass as function of time due to this gravtiational interaction. 

Before we move any forward, it is worth recalling that the inferred LMC's orbital properties, quoted both here and in K13, 
do suffer from a number of simplifications. If the LMC orbits are significantly affected by this simplifications, then the estimated phase-space displacement of the MW center of mass could also be significantly affected.  

On the one hand, as shown in K13, including a simple 
model for the time evolution of the MW potential would increase these orbital periods
by $\lesssim 2$~Gyr (the exact amount depends on the mass
of the host and the satellite). In addition, we have only considered the orbits associated with the 
mean LMC velocity presented in K13. The relatively large uncertainty on each of the velocity 
components, of $\approx \pm 19$ km/s, will yield, in some cases, orbits with
significantly larger periods. Furthermore, we have neglected LMC mass loss due to the 
tidal interaction with the MW. Note, however, that perturbations on the inferred orbital properties 
due to LMC mass loss are not expected to be significant in those cases where the LMC is clearly undergoing its first 
infall. On the other hand, the treatment of dynamical friction implemented here may be overestimated, 
thus artificially increasing the orbital periods (see Section~\ref{sec:lmc_meto}). 

A more accurate 
determination of the LMC's orbital properties as a function of LMC mass 
would require a full $N$-body treatment. Even though this is not the goal of this work, to explore whether 
our approximations regarding dynamical friction and LMC's mass loss are valid, we have ran a full
$N$-body simulations considering one of the MW+LMC mass combinations analyzed in this work. The MW-like host was modeled as a
self-consistent three-component system consisting of a NFW dark matter halo, an exponential stellar disk, and a central bulge following
a Hernquist profile. The LMC galaxy was modeled as a self-consistent Plummer sphere. The masses and parameters that specify each
galactic component were chosen to reproduce the  analytic rigid representation of the galactic potentials associated with a MW  of
$M_{\rm vir} = 1 \times 10^{12}~M_{\odot}$ and a LMC of total mass  $M_{\rm LMC} = 1.8 \times  10^{11}~M_{\odot}$. These parameters are
listed in Table~\ref{tab:N_model}. Initial positions and velocities for the LMC and MW centers of mass were obtained from the
numerically integrated orbits using the analytic rigid potentials. The simulations were started at a lookback time  equal to 2 Gyr. 

\begin{figure}
\centering
\includegraphics[width=80mm,clip]{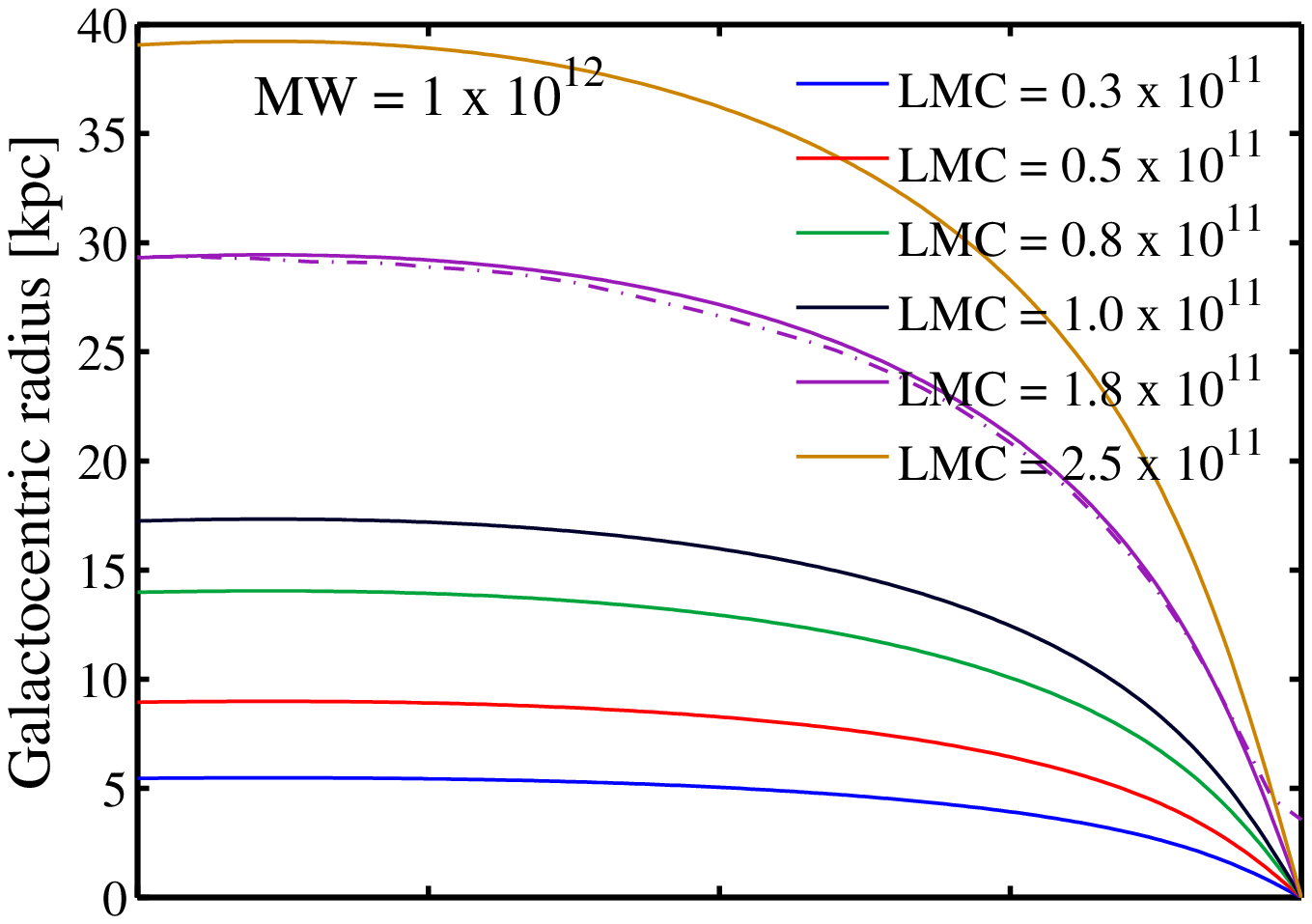} \\
\includegraphics[width=80mm,clip]{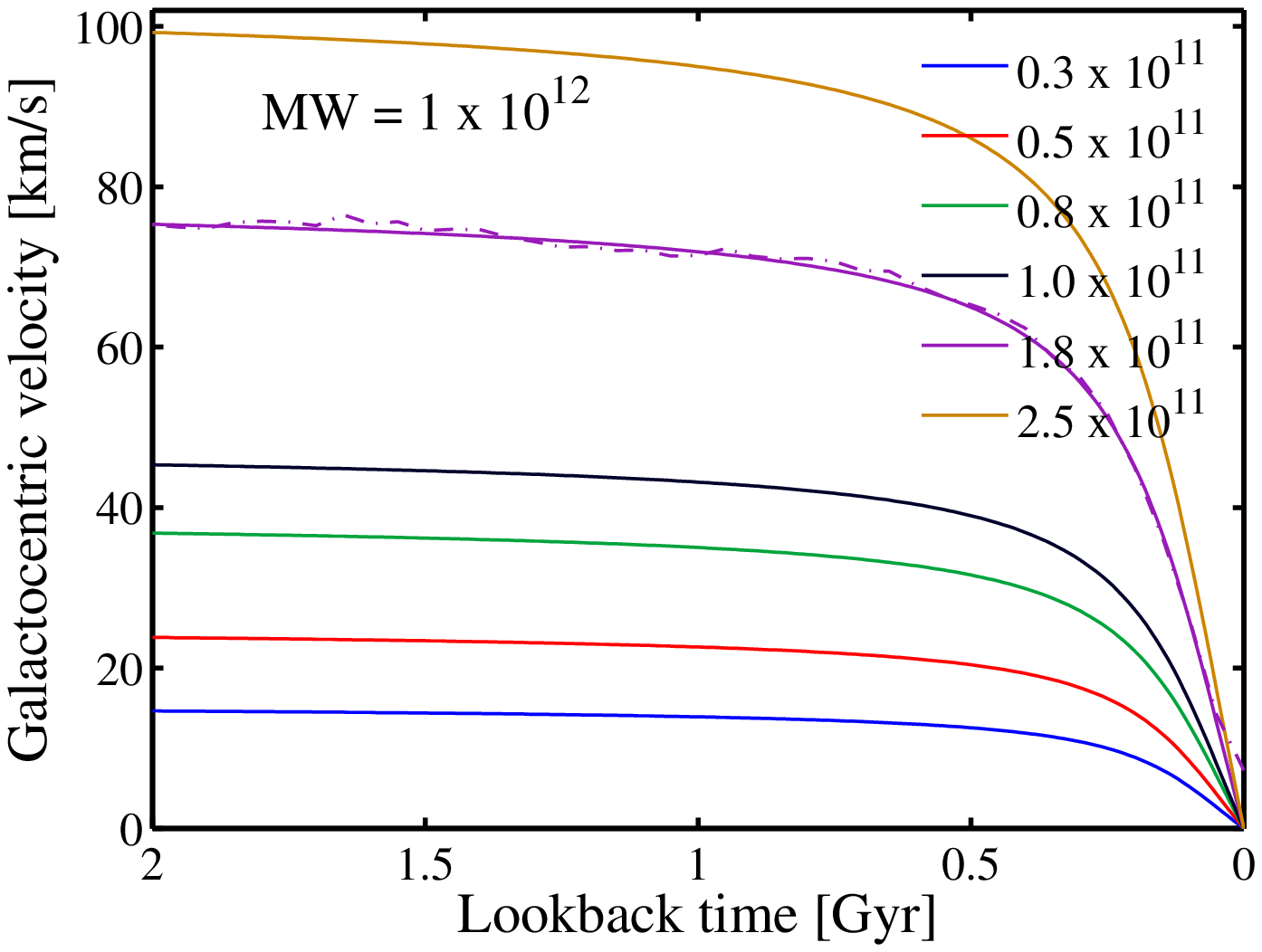}
\caption{Time evolution of the position  and velocity of the MW
  center  of mass  with respect  to its  position at  $t=0$  Gyr.  The
  results are obtained  from simulations where the MW model is
  allowed  to react  to the  gravitational  pull exerted  by the  LMC.
  Orbits  are  integrated backwards  in  time.   Note  that $t=0$  Gyr
  corresponds to  present-day. The solid lines show the results obtained with a
  MW model with a dark matter halo of $M_{\rm vir} = 1 \times 10^{12}~M_{\odot}$.  
  The different colors  indicate the results obtained  with different LMC
  models, as  indicated in  the top right  corner of the top  panel.  The
  most significant  changes in both  the position and the  velocity of
  the MW center of mass  take place only during the last 0.3 to
  0.5 Gyr,  the time at which both, the  MW mass enclosed  within the LMC
  Galactocentric radius becomes comparable to that of the LMC and the distance between both galaxies 
  becomes short enough.  For comparison, the dashed dotted line shows the displacement of the MW center of mass obtained from a fully live $N$-body simulations considering an LMC model with $M_{\rm LMC} = 1.8 \times 10^{11}M_{\odot}$. Note the very good agreement between the results obtained with the rigid analytic potential and their fully live $N$-body counterpart.}
\label{fig:mw_com}
\end{figure}

In Figure~\ref{fig:mw_com} we show, with purple lines, the  time evolution of the position (top panel) and the velocity (bottom panel) 
of the MW center of mass,  with respect  to  its present-day coordinates, due to the gravitational interaction with the LMC model
$M_{\rm LMC} = 1.8\times 10^{11}$ M$_{\odot}$. The solid line show the result from the rigid analytic representation of the 
potentials while the dashed dotted line shows the result from for the fully $N$-body representation.  
Note the very good agreement between the results obtained from the two different modelling techniques. 
The final phase-space displacement of the MW center of mass in the 
$N$-body model is slightly smaller than what was obtained with the analytic rigid case. The differences are  $\approx 4$ kpc and 
$\approx 7$ km/s in position and velocity, respectively. Furthermore, we find the final location of the LMC center of mass in the N-body
model to be ~4 kpc further away from the MW center of mass than its observationally constrained Galactocentric distance. Thus, it is 
likely that by selecting a more suitable set of ICs for the N-body simulations, such that in these calculations the LMC center of mass 
finish at the desired location, this (already small) discrepancy could even become smaller. 

We conclude that our approach of using rigid, spatially extended density distributions to compute the orbital evolution of the 
LMC and MW is an adequate approximation for this study. With this in mind, for the remainder of this section we will only 
consider  rigid analytic potentials. 
In Figure~\ref{fig:mw_com}, different colored lines indicate the time evolution of the position and the velocity 
of the MW center of
mass induced by different LMC-like models. In all cases, the MW model corresponds to that with 
$M_{\rm vir} = 1 \times 10^{12}~M_{\odot}$. Note that, for this low-mass MW model, \emph{the LMC is on its first infall about 
the MW, regardless of the LMC mass}.  

Changes in  both position and velocity  are very rapid and take place
primarily during the last $\approx 0.3-0.5$~Gyr.  As the LMC
approaches its present-day position, $R_{\rm LMC} \approx 50$ kpc, the
mass of the  MW enclosed within a  radius of $R_{\rm LMC}$
becomes smaller. Assuming a MW model with 
$M_{\rm vir} = 1 \times 10^{12}~M_{\odot}$, at present day $M_{\rm MW}^{50}  \approx 3.7 \times
10^{11}~M_{\odot}$ becomes comparable to  the mass of the LMC.  Thus, the
orbital barycenter of  the MW + LMC system  is significantly displaced
from the MW center of mass. For example, for a LMC model with a
total mass  $M_{\rm LMC} = 1.8 \times  10^{11}~M_{\odot}$, the orbital
barycenter is located  at $\approx 14$ kpc from the  MW center of mass
at the present epoch.  In this model, as the LMC approaches its current
location the MW is displaced by $\approx 30$ kpc and its  velocity has 
changed by $\approx 75$ km/s in $\approx 0.5$ Gyr. 

Interestingly, we
find very similar phase-space displacements in our more massive MW models.
The circular velocity at the location of the Sun limits the amount of mass that can exist in the DM halo at small radii. 
As such, to create more massive MW models, the bulk of the mass is added at radii beyond 50 kpc. Correspondingly, the virial radii 
of the halos increase. This means that the resulting mass enclosed within 50 kpc is not strikingly different in the three MW models
we adopt.

The results presented in this section highlight the importance of self-consistent modeling of the MW and LMC 
interaction when trying to constrain the LMC's orbital properties. Given the magnitude of the effect of this 
interaction on the motion of the MW, it is of interest to explore its implications 
for stars in the MW stellar halo. Owing to the extended nature of the stellar halo, not all stars will be 
accelerated at the same rate. Thus, this may affect the observable properties of spatially extended stellar streams.

\begin{figure}
\centering
\includegraphics[width=90mm,clip]{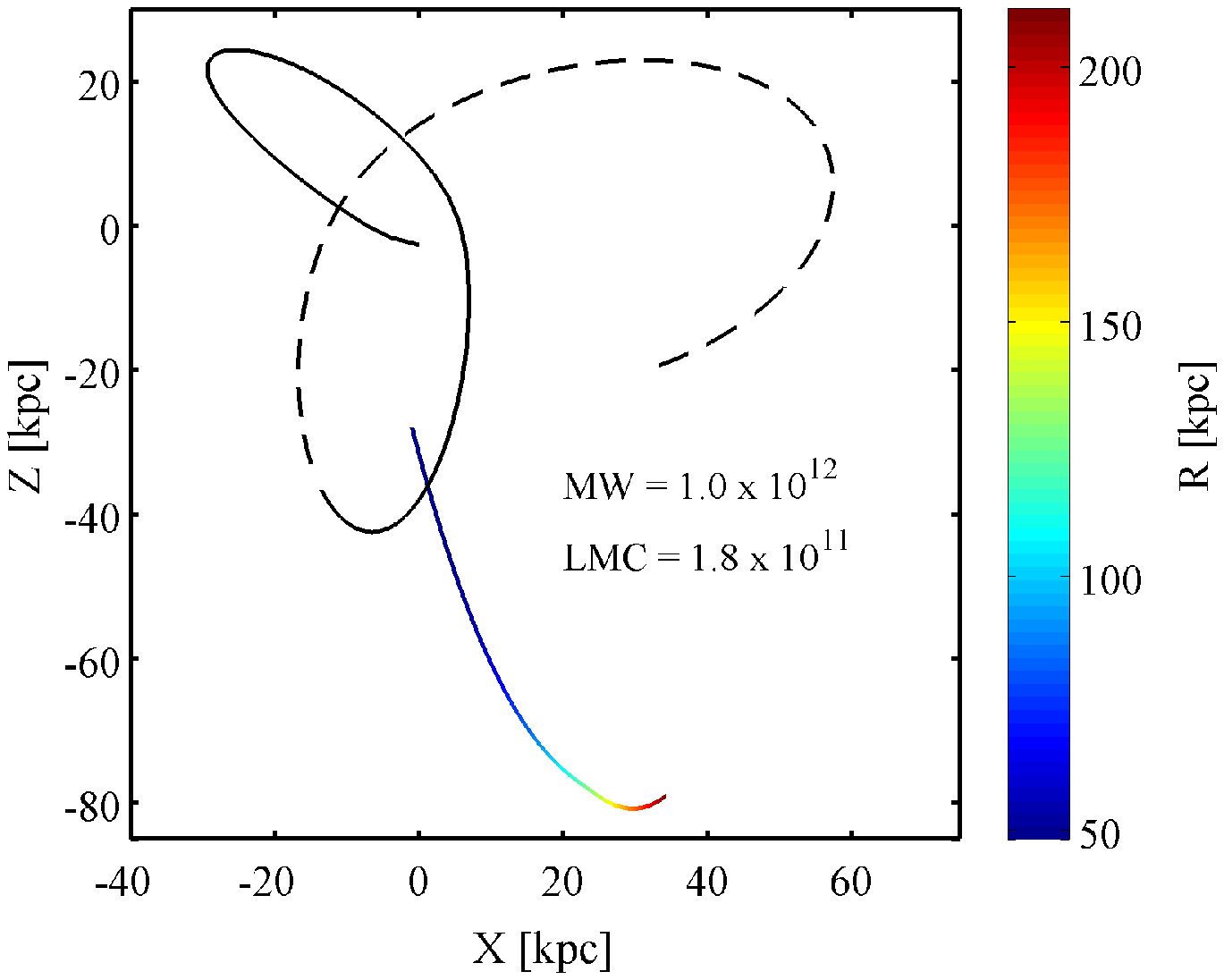}\\
\includegraphics[width=90mm,clip]{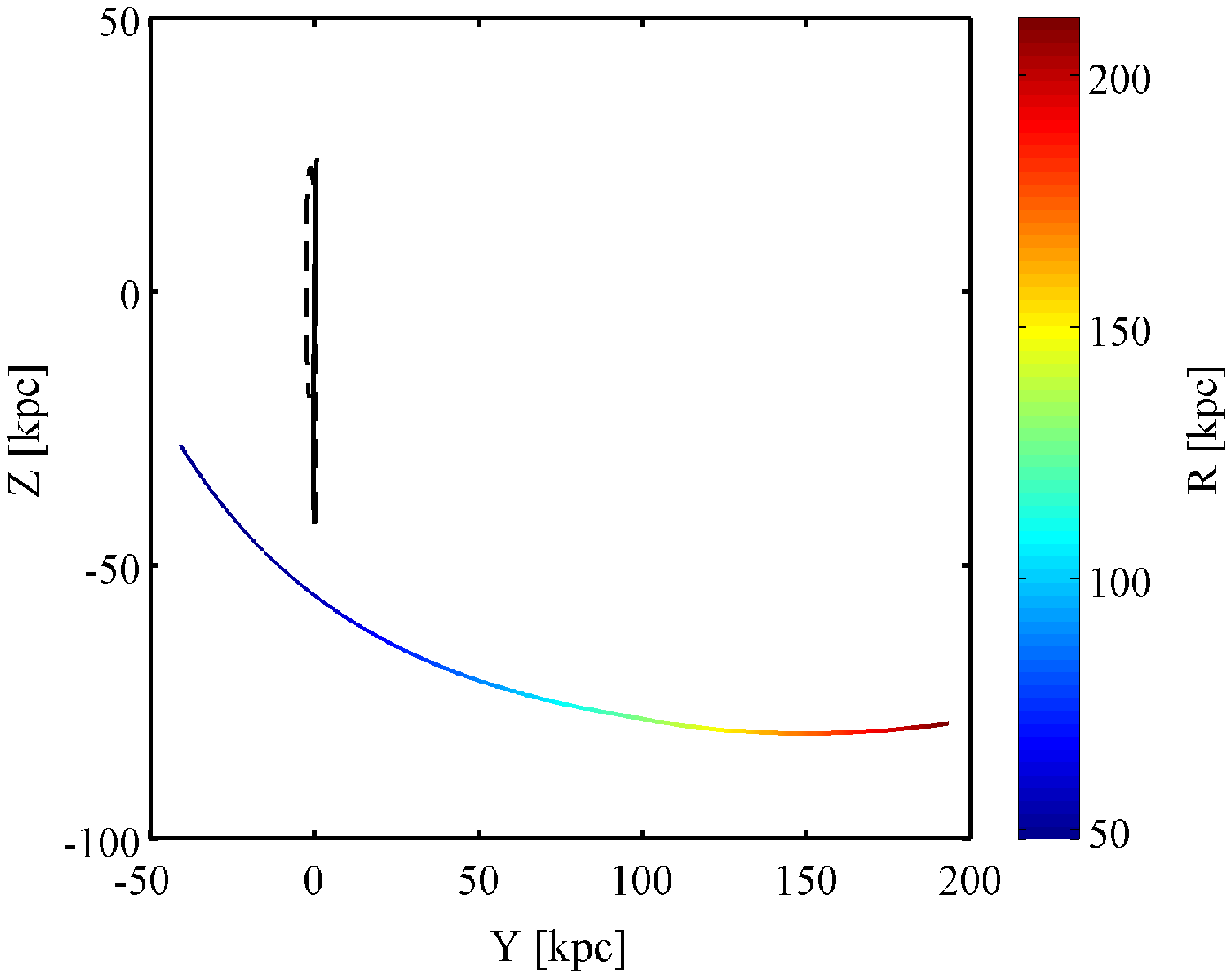}
\caption{Sgr-like  and LMC orbits obtained  in a free MW model with a dark matter  halo of $10^{12}~M_{\odot}$. 
For this example an LMC model with total mass $M_{\rm LMC} =  1.8 \times 10^{11} M_{\odot}$  was considered.
The orbits are illustrated 
in the $XZ$ (top panel) and $YZ$ (botom panel) Galactocentric planes, with $Z$ pointing towards the Galactic pole and $X$ pointing 
  towards the opposite direction of the Sun. The black and color coded lines indicate the Sgr-like and the LMC orbit, 
  respectively. The color coding indicates the LMC's galactocentric distance.
  Solid and  dashed black lines show the Sgr's first and  the second Gyr of backwards evolution, respectively. For clarity, the LMC orbit
  is only shown during the first Gyr of backwards evolution.}
\label{fig:lmc_sgr_orb}
\end{figure}

\begin{figure*}
\centering
\includegraphics[width=180mm,clip]{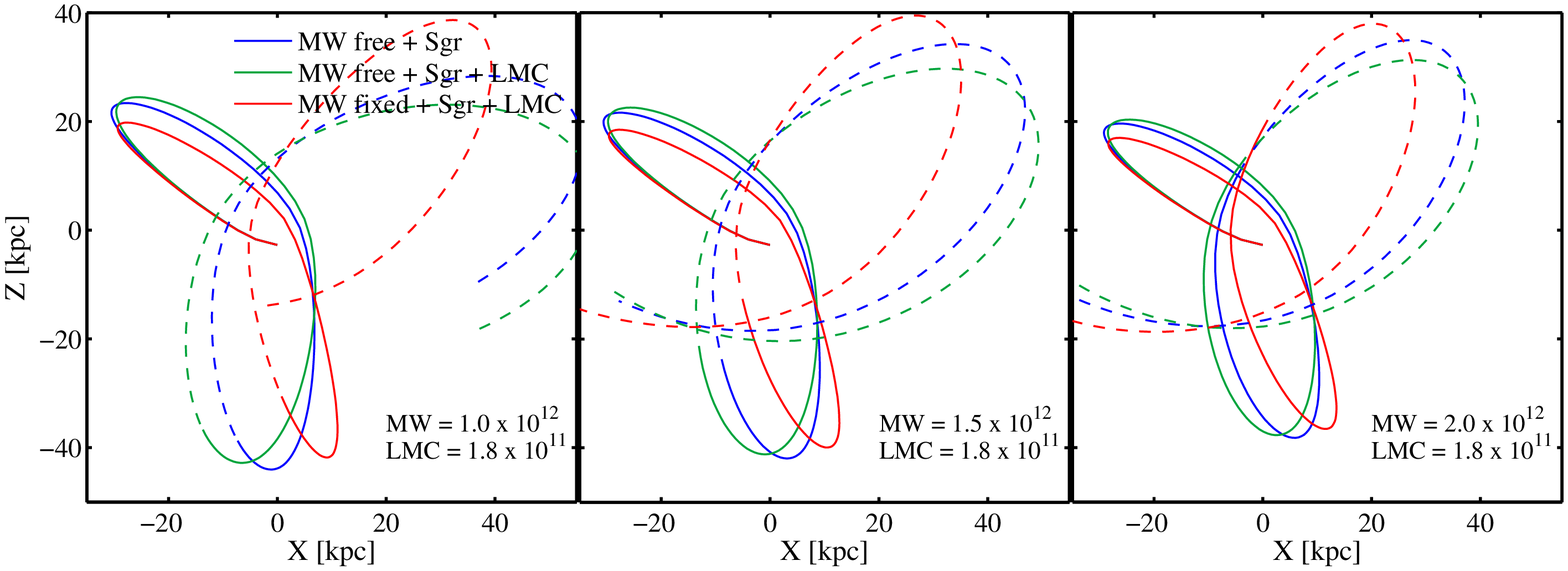}
\caption{The  different  lines  show  Sgr-like  orbits  integrated for 2 Gyr in
  different models of the  MW potential. The orbits are illustrated
  in the $XZ$ Galactocentric plane, with $Z$ pointing towards the Galactic pole and $X$ pointing 
  towards the opposite direction of the Sun. The LMC moves in a direction that is
  approximately perpendicular to this plane. Sgr orbits are integrated
  backwards in time.  Thus, present-day positions are the same in all
  cases. Solid and  dashed lines show the first and  the second Gyr of
  backwards evolution, respectively. The  blue lines show the Sgr-like
  orbits obtained in  MW models that are  allowed to react to
  any external perturbation (Sgr + MW only). The green  lines show the same orbits, now
  introducing a LMC model with a mass of $M_{\rm LMC} =
  1.8 \times  10^{11}~M_{\odot}$, which follows the orbit described in 
  Figure~\ref{fig:lmcs}. The red  lines show  the corresponding Sgr orbit
  obtained after  artificially fixing  the MW  center of
  mass, but including the perturbative effects of the LMC.  
  From  left to  right, the different  panels show  the results
  obtained using MW models with masses $M_{\rm vir}  = 1$, 1.5 and
  $2   \times   10^{12}~M_{\odot}$,   respectively.  Note   that   the
  differences on the Sgr-like orbits between ``free'' (green line) and
  ``fixed'' (red line)  MW  models are  even larger  that those
  obtained in models with and without the LMC (green and blue lines).}
\label{fig:orbits}
\end{figure*}

\section{The orbit and tidal debris from Sgr}
\label{sec:sgr}

In this section we explore the implications of the motion of the MW around its center of mass 
with the LMC for the orbit of the Sagittarius dwarf galaxy (Sgr) and its distribution of tidal 
debris. For this purpose,  we will integrate Sgr-like orbits in different 
scenarios. We will consider spatially ``free" and 
``fixed" MW-like hosts, with and without the presence of a LMC-like satellite.
As in Section~\ref{sec:lmc}, the experiments analyzed in this section will assume  smooth and analytic 
representations of the  Galactic potentials. The models and methodology 
are described in Section~\ref{sec:analy_sgr_mod}. We briefly review 
the main properties of the Sgr stream and discuss 
previous attempts to constrain the shape of the MW dark matter halo based on the stream's phase-space distribution  
in Section~\ref{sec:sgr_nature}. Our results are presented on Section~\ref{sec:ana_sgr}. 
Note that, throughout this study, we are  \emph{not}  
interested in obtaining an orbit that could accurately reproduce the observed  
distribution of the Sgr tidal debris. Our  goal is rather to simply characterize  the  significance
of  artificially fixing the  MW center of mass on Sgr-like orbits.  
If the effect is significant, this justifies the incorporation of complete and realistic modeling 
of the LMC+MW interaction in future analyses of Sgr and other long stellar streams in the halo of the MW.

\subsection{The  complex  nature   of  the  Sgr  streams}
\label{sec:sgr_nature}

As  discussed in  Section~\ref{sec:intro},  the Sgr  tidal
stream and its  remnant core have been used multiple  times in the past
to  probe the mass  distribution of  the MW.  The  main reason
behind the wide popularity of  this satellite galaxy is the very large
radial  and angular  extent covered  by its  debris. The  Sgr stellar
stream     spans    at least $300^{\circ}$    across     the     sky
\citep{1997AJ....113..634I}, and observations suggest  that it
can  be observed  at Galactocentric  distances as  large as  $100$ kpc
\citep[e.g][]{2003ApJ...599.1082M,2003ApJ...596L.191N,2011ApJ...731..119R,
2013ApJ...765..154D}. As discussed by \citet{2014MNRAS.439.2678D},  
the Sgr  stream has  a  very complicated
structure, making it difficult to model. The mean orbital poles of
the great  circles that best fit  debris leading and  trailing the Sgr
core      show     a      difference     of      $\sim     10^{\circ}$
\citep{2005ApJ...619..800J}.  Stars  in the trailing  and leading arms
show   very  different   apocenters   \citep{2014MNRAS.437..116B}  and
bifurcations     have      been     observed     in      both     arms
\citep{2006ApJ...642L.137B, 2012ApJ...750...80K, 2013ApJ...762....6S}.
  In addition, \citet{2010MNRAS.408L..26P} showed that the phase-space
  configuration of the Sgr stream strongly depends on the structure of
  the progenitor.

  Given the complex  nature of this stream, it  is not surprising that
  several  previous studies  have yielded  contradictory  results with
  regards to  the structure of the MW's gravitational potential.
  For example, the previously mentioned  tilt of the orbital plane can
  be reproduced with $N$-body simulations if the MW dark matter
  halo     is    modeled    as     a    mildly     oblate  galactic  component
  \citep{2005ApJ...619..800J}.   On the  other  hand, radial  velocity
  measurements of a sample of M  giant stars favor a prolate dark matter
  halo             \citep{2004ApJ...610L..97H}.              Furthermore,
  \citet{2010ApJ...714..229L} showed that  a triaxial dark matter halo
  model could reproduce the angular position, distance, and
  radial velocity constraints imposed by current wide-field surveys of
  the Sgr stream. However, the results from this model are bound by a 
  number of caveats.   Firstly, the  model requires  the disk's  minor axis  to be
  aligned with the  intermediate axis of the triaxial  halo.  As shown
  by  \citet{2013MNRAS.434.2971D},  this  configuration  is  extremely
  unstable.   The problem  can be  alleviated if  the assumption  of a
  disk-halo  alignment is relaxed  when searching  for a  best fitting
  Galactic   potential  \citep{2014MNRAS.439.2678D}.    Secondly,  the
  resulting axis  ratios are not compatible  with expectations derived
  from   cosmological    simulations   \citep[e.g.,   see   discussion
  by][hereafter   VCH13]{2013ApJ...773L...4V}.    Interestingly,   VCH13
  showed  that if the  gravitational field  of the  LMC is  taken into
  account  when  computing the  orbit  of   Sgr,  the
  triaxial configuration of the MW-like dark matter halo can be
  brought to a more cosmologically plausible shape.

The analysis presented in VCH13 shows that the torque on Sgr exerted by
the LMC can be as important as that of the MW's dark
matter halo, introducing non-negligible  perturbations to the orbit of
Sgr  and its  distribution of  debris. Attempts  to reproduce  the Sgr
stream  without a  model  for  the LMC  perturbation  will consequently  force
searches of  the best fitting  parameters that characterize  the MW's 
gravitational potential to artificially adjust in order to account
for this perturbation. 

While a very relevant conclusion, the work of VCH13 (as well as many of
the previously cited works) considered MW models that are fixed
in  phase-space.   In  addition,  their  results were  based  on  test
particle simulations  in which the Sgr stream  is being significantly
perturbed   by  the   LMC  over 3   to   4  Gyr.    As  shown   in
Section~\ref{sec:lmc},  even   relatively  low-mass  first-infall  LMC
models can significantly  accelerate the MW inner  regions in a
very short period of time. It is thus likely that, in a Galactocentric
reference  frame (as opposed  to a  barycentric reference  frame), the
distribution of Sgr  debris, which covers a radial  extension of $\sim
100$  kpc, will  be  significantly perturbed  due  to the  phase-space
displacement of the  MW center of mass.  To  explore this, we \
integrate Sgr-like orbits in MW potentials that are allowed to
freely react to the gravitational pull of the LMC.

\subsection{Analytic models}
\label{sec:analy_sgr_mod}

\subsubsection{Methodology}

As in Section~\ref{sec:lmc}, to follow the gravitational interaction between the
MW, the LMC, and Sgr,  we  used a  symplectic
leapfrog integration scheme \citep{2001NewA....6...79S}. The host
and  the two satellites are  represented  with  analytic potentials;  the
center  of  each  one  follows  the orbit  that  results  from  the
acceleration of the  other two.  In all cases the  orbits are integrated
backwards in time from their present-day positions and velocities.
For simplicity, in this section we model the dark
matter halo of all galaxies with Hernquist profiles. This allows us to model 
the dynamical friction that each of our three galaxies (LMC, Sgr, and 
MW) induces on the remaining two by approximating the integral in Equation~\ref{eq:chandra} as follows:
\begin{equation}
\label{eq:chandra_ap2}
\int_{0}^{v_{2}} v^2
f(v){\rm d}v \approx \frac{1}{6}\left( {\rm erf}(x)- \frac{2x}{\sqrt{\pi}} {\rm e}^{-x^2} \right),
\end{equation}
where $x  = 2 v_{2} \sqrt{r_{1}/(M_{1}G)}$.  Here, $r_{1}$ is the scale radius of the galaxy 
causing the friction. The Coulomb factor 
is computed as 
\begin{equation}
\Lambda = \frac{rv_{2}^{2}}{G M_{2}(r)},
\end{equation}
where $r$ is the distance between the centers of the two galaxies and $M_{2}(r)$ is 
the mass of the galaxy being decelerated enclosed within $r$ \citep{2013MNRAS.435L..63C}.

\begin{figure}
\centering
\includegraphics[width=80mm,clip]{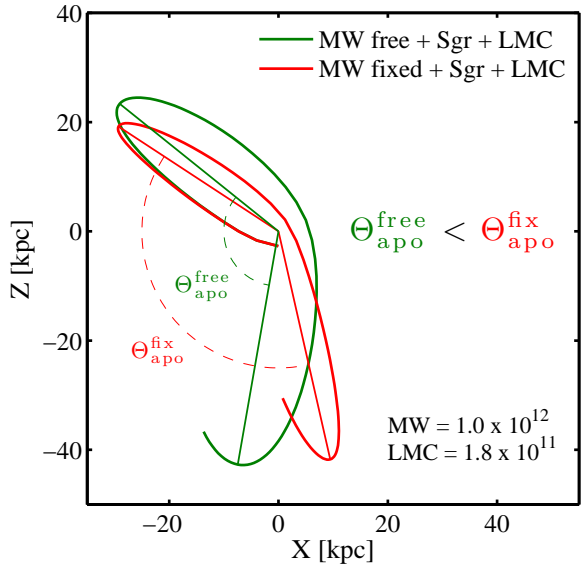}
\caption{Angular distance between two consecutive apocenters obtained from Sgr-like orbits 
integrated in a free (green) and a fixed (red) MW model. In both cases, a model for the LMC 
is also included in the simulations. For clarity, orbits are shown only during the first Gyr 
of backwards evolution. Note that  allowing the MW model 
to react to the pull  of its satellites, especially the LMC, results in a 
significant decrease of $\Theta_{\rm apo}$.}
\label{fig:theta_apo}
\end{figure}

\begin{figure*}
\centering
\includegraphics[width=180mm,clip]{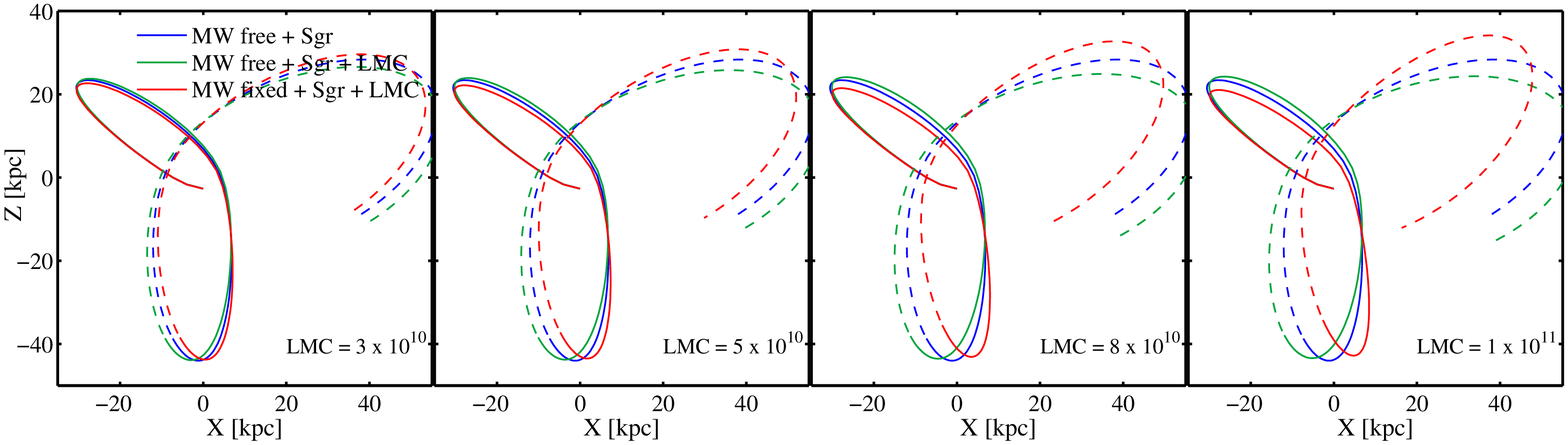}
\caption{As in  Figure~\ref{fig:orbits}, but for LMC  models with different
  total  masses. In  each panel,  the  mass of  the corresponding  LMC
  model is indicated in the bottom right. In all cases, we consider a
  MW potential   with   mass  $M_{\rm   vir}   =  1   \times
  10^{12}~M_{\odot}$. Note that perturbations in the Sgr-like orbits are
  noticeable in all simulations that include LMC models with masses $> 5 \times
  10^{10}~M_{\odot}$.}
\label{fig:diff_lmc}
\end{figure*}

\subsubsection{Galactic models}
\label{sec:analy_sgr_pot}

We model the MW as a three component system. The main difference 
with the MW model presented in Section~\ref{sec:pot_gal_smooth} is the profile of its 
DM halo. In this Section, all DM halos follow
a Hernquist profile (Eqn.~\ref{eq:hernq}). The parameters that describe our MW models are listed in Table~\ref{tab:mw}. 
The scale radius of the Hernquist profile DM halos, $r_{H}$, are obtained from the NFW scale radii 
listed in Table~\ref{tab:mw}  following
\citet{2005MNRAS.361..776S},
\begin{equation}
\label{eq:nfwtohern}
r_{\rm H} = r_{\rm s} \sqrt{2 \left( \log(1+c) - \frac{c}{1+c} \right)}.
\end{equation}

To  smoothly model  Sgr,  we also choose  a
Hernquist  profile.   The  model   parameters  are based on those presented by
\citet[][hereafter P11]{2011Natur.477..301P}. Our single component model consists of a
DM halo with a mass, prior to crossing  the MW
virial    radius    ($R_{\rm     vir}$),    of    $M_{\rm    Sgr}    =
10^{11}~M_{\sun}$. As described by P11, this large value of $M_{\rm Sgr}$ prior to infall 
is obtained from a cosmological abundance matching argument 
\citep{2009ApJ...696..620C,2010ApJ...717..379B}, based on the present-day luminosity of the 
Sgr core and tidal debris. Lower mass models of Sgr are presented  in Section~\ref{sec:nbody_sgr}.
Note however that, as in P11 and references therein, the satellite 
is initially launched 2 Gyr ago at a distance of 80 kpc from the Galactic centre, traveling vertically at 
80 km s$^{-1}$ toward the North Galactic Pole.  Thus its mass, at this point in time, is
truncated at  the instantaneous Jacobi  radius $r_{\rm J}  \approx 30$
kpc. This  leaves a total bound mass (2 Gyr ago)  of $M_{\rm Sgr}^{80} \approx  3.8 \times
10^{10}~M_{\odot}$,  i.e.,  a  factor of  $\sim 3$  smaller  than its
effective virial mass at infall, $M_{\rm vir}$.  The scale length of the profile is
$R_{\rm  Sgr} = 13$  kpc.\footnote{Note that  the equivalent  NFW scale
  radius is 6.5 kpc (see Eq.~\ref{eq:nfwtohern}).} In order to crudely account
for  Sgr galaxy's mass loss due  to tidal interaction  with the MW
potential, we assume that its  mass linearly varies
during the 2 Gyr of evolution between $M_{\rm Sgr}^{80}$ and 
$M_{\rm Sgr} = 10^{9}~M_{\odot}$ \citep{2005ApJ...619..807L,2011Natur.477..301P}.
A Hernquist profile is also used  to model the LMC. The parameters that specify each of our LMC models
are listed in Table~\ref{tab:lmc}.

\subsection{An analytic treatment of the MW + LMC + Sgr system}
\label{sec:ana_sgr}

In this Section we characterize the significance of the perturbative effects associated 
with a first-infall LMC on the orbit of Sgr. In all  experiments, the orbits of 
our three galaxies are integrated backward in time 
 for 2 Gyr from  their present-day positions. To be
consistent with the full $N$-body  integrations that we analyze  in
Section~\ref{sec:nbody_sgr},  present-day initial conditions for  our Sgr-like orbits
are  obtained as  follows. As discussed in Section~\ref{sec:analy_sgr_pot}, we first integrate our Sgr model 
forward in time for 2 Gyr in a free MW model. The initial conditions at this initial time are  
$(X,Y,Z)_{\rm  Sgr} =  (  80, ~0,  ~0)$ kpc  and $(v_{x}, ~v_{y}, ~v_{z})_{\rm Sgr} = (0,~0,~80)$ km/s. 
Note that, only in this first step, we neglect Sgr's mass loss due to the tidal interaction with 
the MW potential. The position and velocity of the Sgr model at the final integration point 
(i.e., after 2 Gyr of evolution), $(X,Y,Z)_{\rm  Sgr} =  (  0, ~0,  ~-3)$ kpc  and
$(v_{x}, ~v_{y}, ~v_{z})_{\rm Sgr} = (413,~0,~-46)$ km/s,  are then  used  as present-day initial 
conditions for the backward integration. As an example, we show in Figure~\ref{fig:lmc_sgr_orb} the resulting Sgr-like 
and LMC orbits obtained  in a free MW model with a dark matter  halo of $10^{12}~M_{\odot}$. 
For this integration an LMC model with total mass $M_{\rm LMC} =  1.8 \times 10^{11} M_{\odot}$  was considered.

In Figure~\ref{fig:orbits}, we compare the resulting backward integrated Sgr-like orbits obtained 
with free and fixed MW models, with and without the LMC. The blue
line in the left panel shows the backward time
integrated Galactocentric orbit of our Sgr-like satellite  in a MW model with a
dark matter  halo of $10^{12}~M_{\odot}$.  The solid  and dashed lines
indicate the first and second Gyr of evolution, respectively.  In this
orbital integration, the MW  center of mass is allowed to react
to any  external potential. However, the  initial mass of  Sgr, $10^{9}~M_{\odot}$, 
is very small and thus the orbital  barycenter is approximately located at the
MW center of mass.

The  green line in the same panel shows the
orbit  of Sgr  in a  Galactocentric reference  frame, now  including a
model for the LMC. As before, we  allow the MW center of mass to react
to the  pull of any external  potential.  For this  experiment we have
chosen a  LMC model with  a total mass  of $M_{\rm LMC} =  1.8 \times
10^{11} M_{\odot}$ that is  experiencing its first pericentric passage
at the     present-day      (see      Section~\ref{sec:lmc}      and
Table~\ref{tab:lmc}).  This LMC  model represents  the  canonical model
described by  K13.  Such a massive LMC  is required to  keep the LMC-SMC
binary configuration  for longer than 2  Gyr in MW models with
masses $\leq 1.5 \times 10^{12}~M_{\odot}$ \citep{2010ApJ...720L.108G,
  2013ApJ...768..140B, K13, 2014A&A...562A..91P}. It has also been used in the past 
  to successfully reproduce many of the observable properties of the 
  Magellanic stream \citep{2012MNRAS.421.2109B}.  A comparison of the
blue and green lines shows that  the perturbation on the Sgr orbit due
to  the  LMC's gravitational  pull  is  indeed  significant, as  first
suggested by VCH13. Note, however, that  in our case we are considering a
first infall scenario  for the LMC, and so its perturbative effects
have only operated over the past $\sim$1.5 Gyr. 

As indicated  by the solid lines,
the  Sgr orbital  perturbation  is  significant even within the past 1 Gyr  of
backwards evolution.  With a red line we now show the orbit of our Sgr
model  in a  Galactocentric reference  frame, including  the  same LMC
model  but, as in VCH13,  keeping  the  MW  center of  mass  fixed  at  all
times. Note the very different orbit for Sgr that is obtained when the
MW is  not allowed to react to the  gravitational pull exerted by
its satellites, especially the LMC.   The differences in  the Sgr-like orbits 
between  a ``free'' (green  line) and  a ``fixed''  (red line)  MW  model  are even
larger that those  obtained in models with (green)  and without (blue)
the  LMC.  

Another noticeable effect is the  different angular  distances 
between the  last two apocenters  of the green and red orbits,  $\Theta_{\rm apo}$.
As shown in Figure~\ref{fig:theta_apo}, allowing the MW model 
to react to the pull  of its satellites, especially the LMC, results
in a significant decrease of $\Theta_{\rm apo}$. A comparison between
the Sgr-like orbits in the ``free'' and the ``fixed'' MW models
that include the LMC yields  a  $\Delta\Theta_{\rm  apo}  \approx 17^{\circ}$. Instead, as can be seen in the left
panel of Figure~\ref{fig:orbits}, a comparison
between the Sgr orbits obtained in free MW models with (green line) and without the LMC (blue line) yields 
a smaller but still noticeable $\Delta\Theta_{\rm  apo}  \approx 8^{\circ}$. 
\citet[][hereafter  B14]{2014MNRAS.437..116B}   finds  a  $\Theta_{\rm
  apo}$ between  the apocenters  of the Sgr  leading and  trailing arms
that is $\sim 25^{\circ}$ smaller than what is predicted for Sgr orbits in logarithmic fixed
halos. Thus, taking  into account a free MW and a model of the
LMC  could at  least  partially explain  this  observed smaller-than-predicted  angular
distance between the consecutive apocenters. Note that the magnitude of $\Delta\Theta_{\rm  apo}$ strongly
depends on the initial orbital conditions of Sgr, as well as on the mass of the three galaxies involved.  
The middle and right panels of Figure~\ref{fig:orbits} show
the same experiments, now in MW models with $1.5$ and $2 \times
10^{12}~M_{\odot}$, respectively. Even in these more massive
MW models the perturbation  to the orbit of Sgr associated with
fixing  the MW center  of mass  is very  significant,  and again
larger than that  obtained by the inclusion of the LMC torques on 
the Sgr orbit alone. 

In Figure~\ref{fig:diff_lmc}  we explore the Sgr orbital history 
including LMC  models of different
total masses. In  all cases, the MW model  contains a dark matter
halo of $1 \times 10^{12}~M_{\odot}$. The left panel shows the results
obtained  with our least  massive LMC  model, $M_{\rm  LMC} =  3 \times
10^{10}~M_{\odot}$. In  this case, perturbations to  the
orbit of  Sgr are  almost negligible, regardless of whether we  include a
model for the LMC or consider a free MW.  However, as we
increase the  mass of the  LMC, the perturbation  on the orbit  of Sgr
quickly  becomes noticeable.  For a  LMC with  $M_{\rm LMC}  =  8 \times
10^{10} M_{\odot}$ (i.e., the mass used by VCH13), the perturbation is very clear.
As before, the largest  changes in the orbital path  of Sgr are
obtained  when   we   allow  the   MW  to  react   to the external
gravitational potential of the LMC.

\begin{figure*}
\centering
\includegraphics[width=85mm,clip]{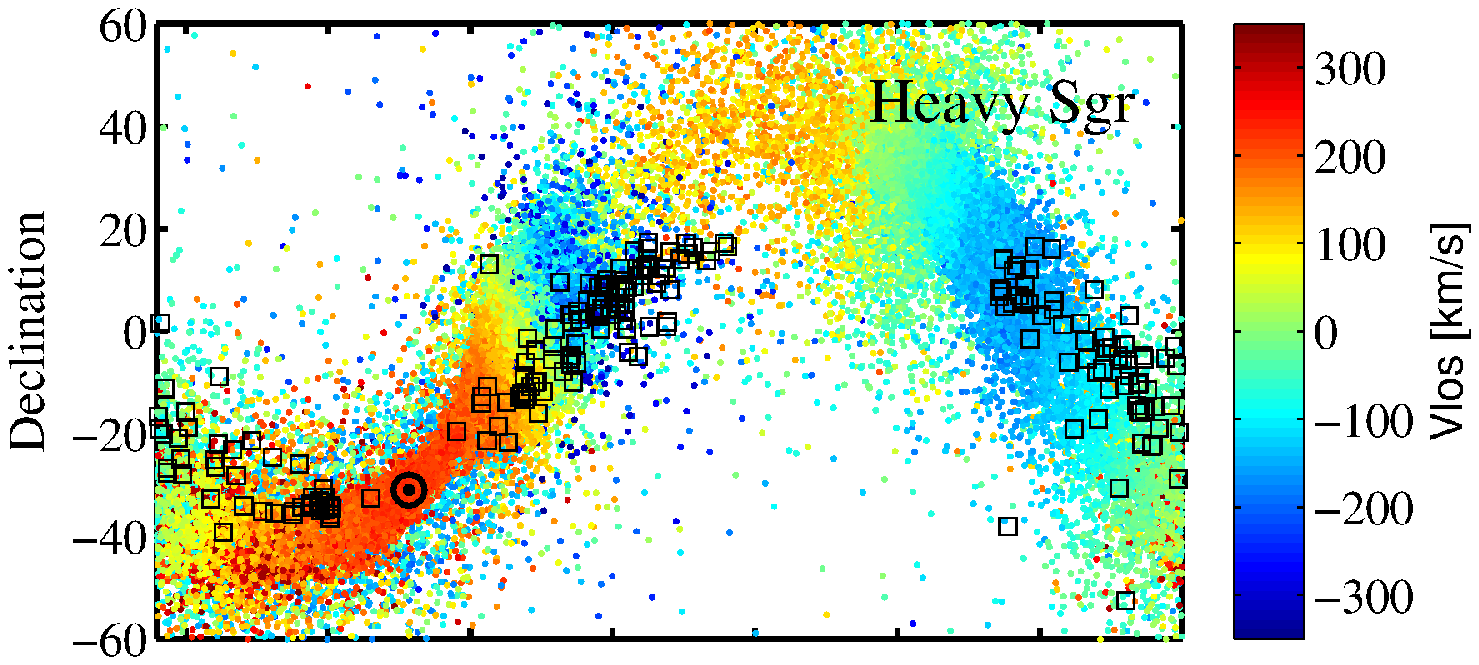}
\includegraphics[width=85mm,clip]{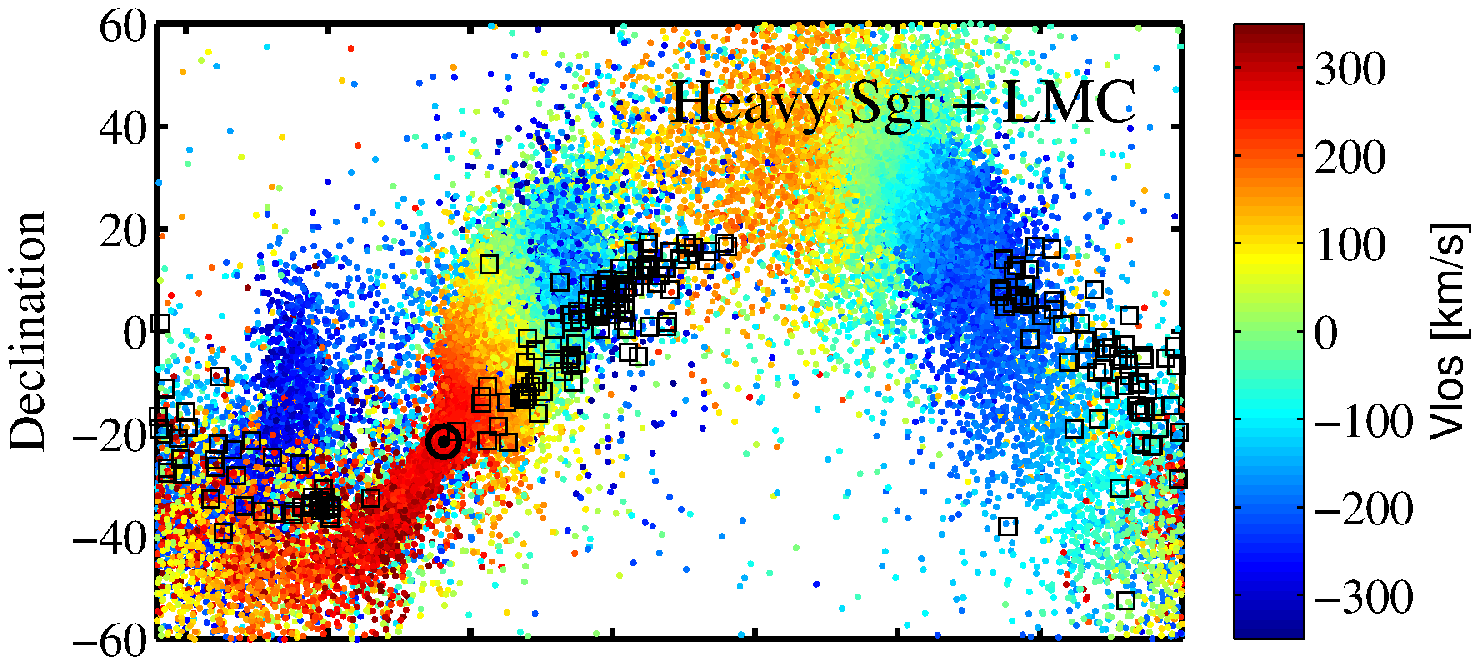} \\
\includegraphics[width=85mm,clip]{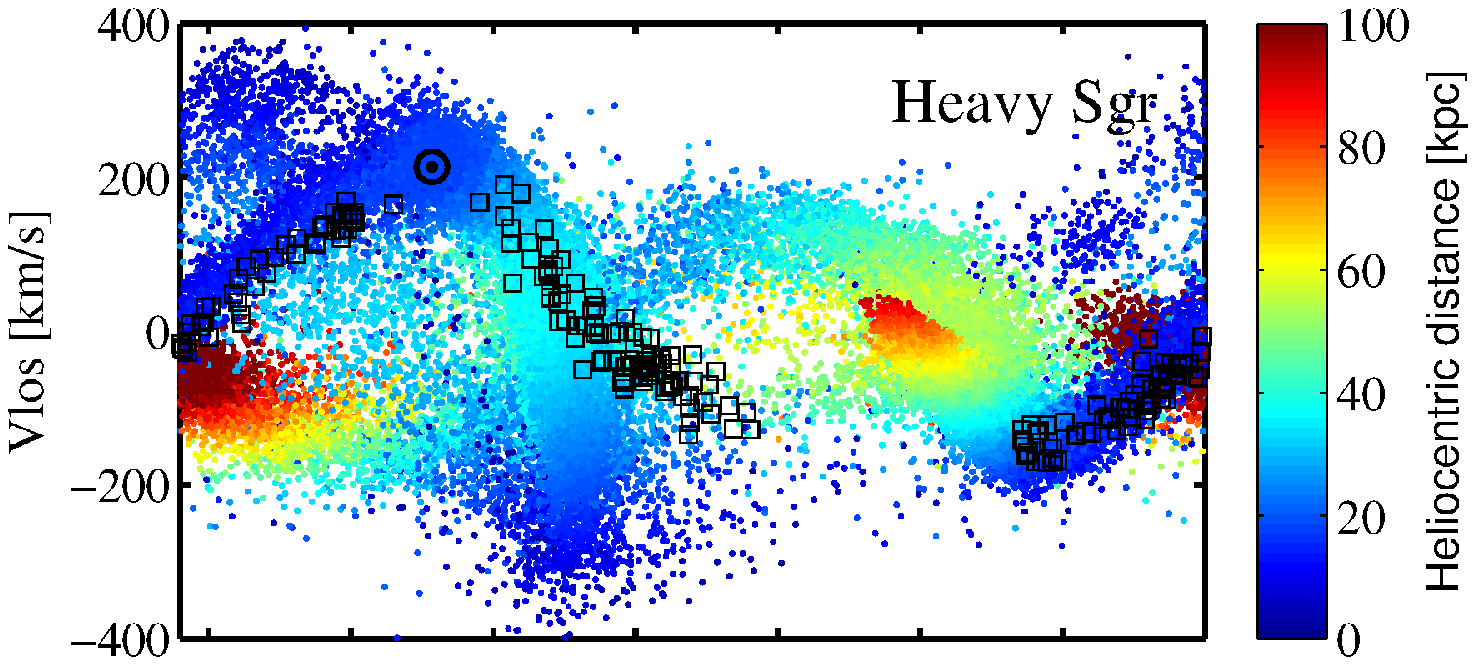}
\includegraphics[width=85mm,clip]{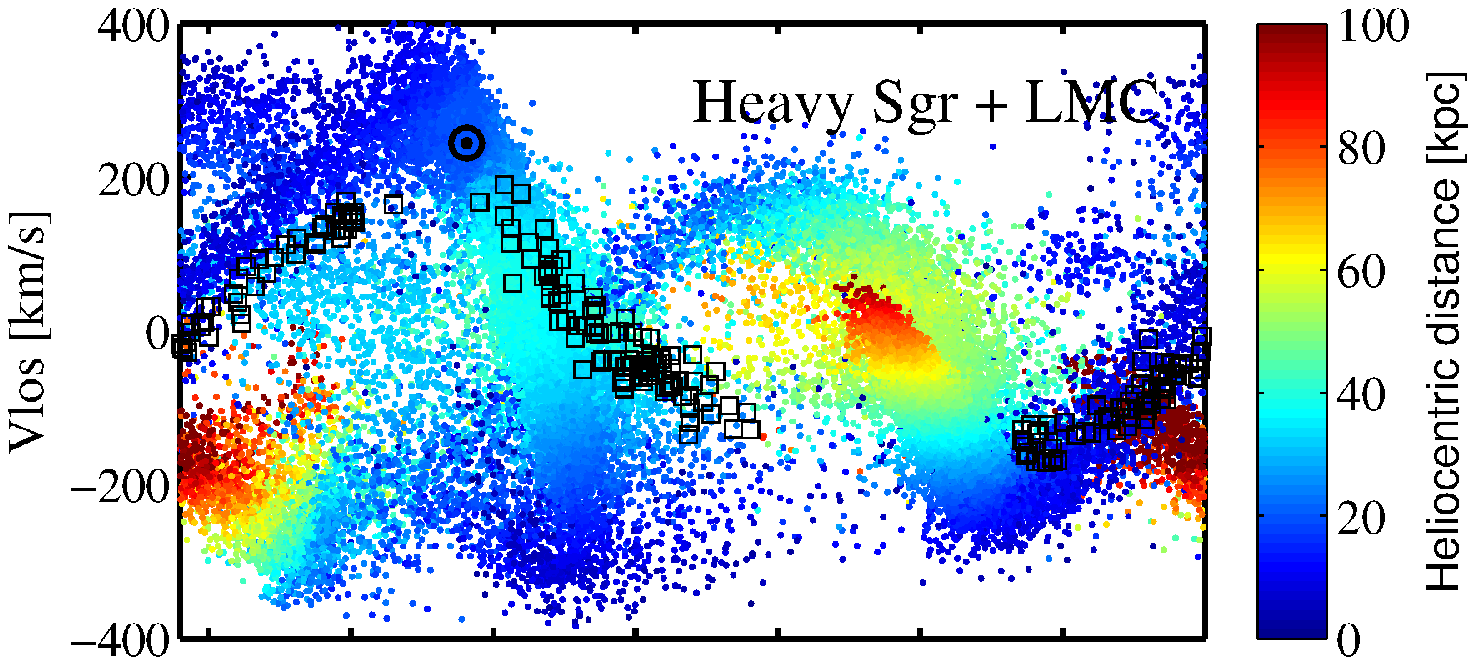} \\
\includegraphics[width=85mm,clip]{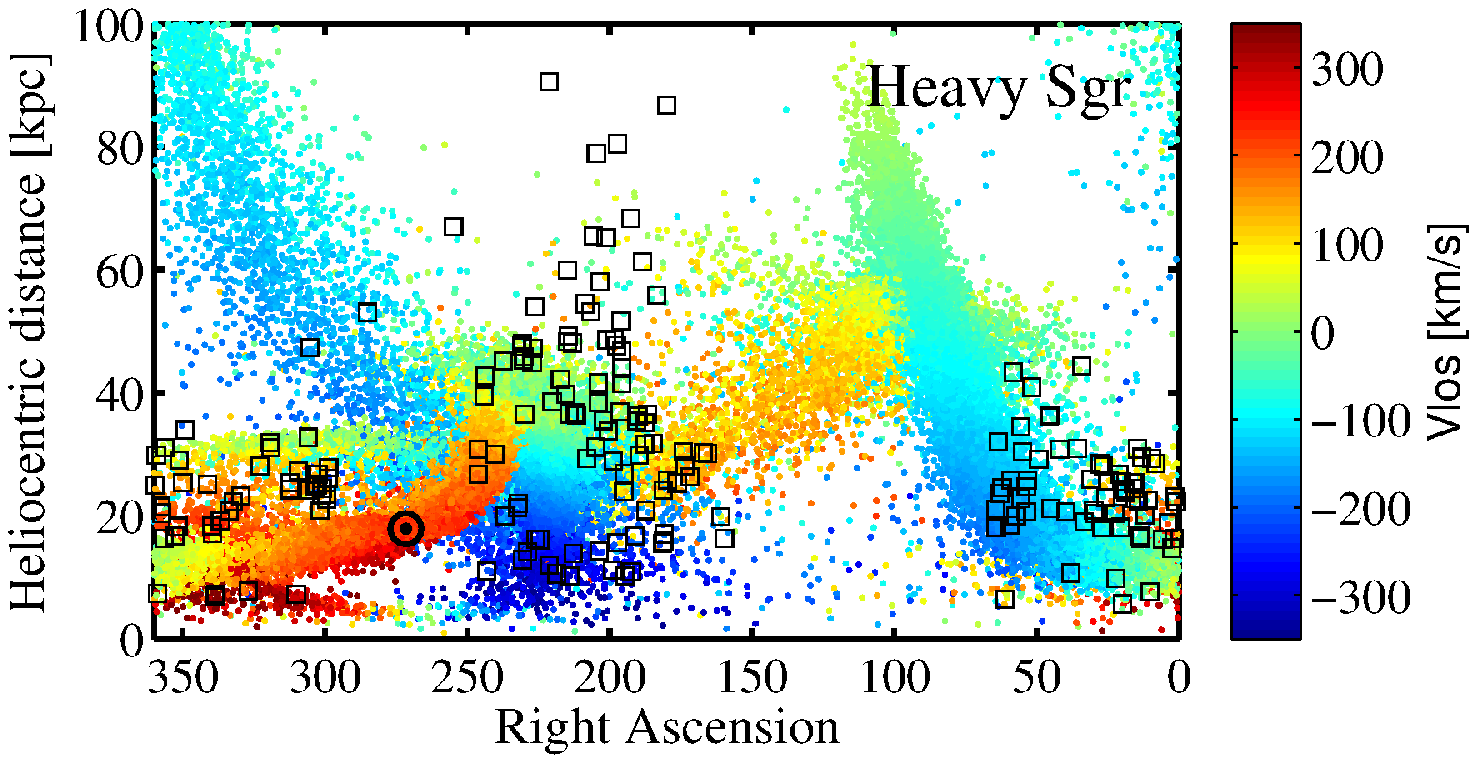}
\includegraphics[width=85mm,clip]{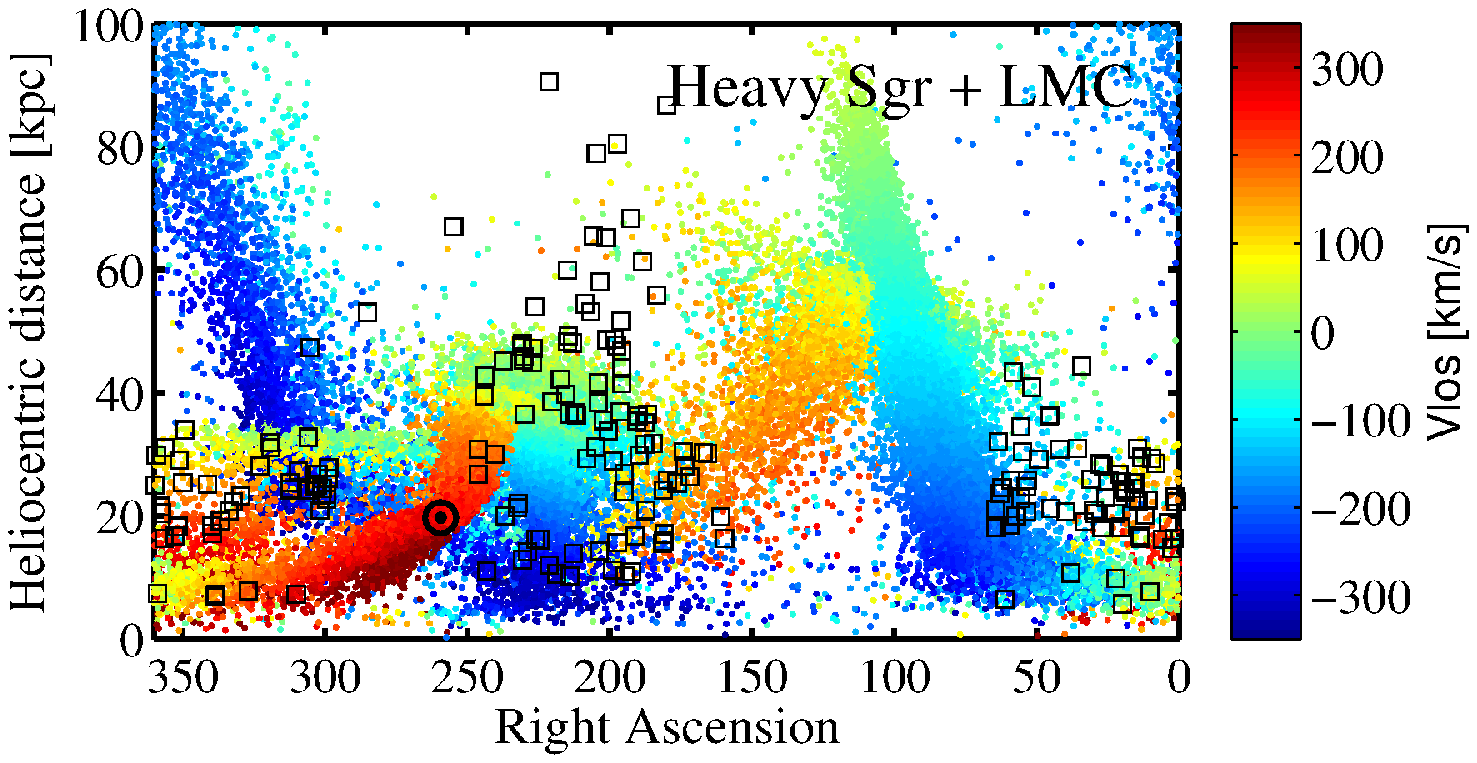}
\caption{Present-day  distribution of  the simulated  Sgr  stream  and core  in
  different  projections  of   phase-space.   These  distributions  are
  obtained from the  simulations with the Heavy Sgr  model. The black circle indicates
  the current location of the simulated Sgr remnant core. From top to 
  bottom we show the stellar particle distribution projected in right ascension
  versus   declination,   right  ascension   versus line-of-sight
  velocity  with respect to the Galactic
  standard-of-rest  (V$_{\rm  los}$), and right
  ascension versus heliocentric distance. The star particles are color
  coded according  to the  quantity indicated in  the color  bars. The
  panels on  the left show  the results obtained after  simulating the
  MW-Sgr interaction  in isolation. The  black squares   show data
  from 2MASS M-giant  stars \citep{2004AJ....128..245M}. The panels on
  the  right  show  the   results  obtained  after  including  in  the
  simulation a  LMC model with total  mass $M_{\rm LMC}  = 1.8 \times
  10^{11}~M_{\odot}$.   Note  that  significant perturbations  to  the
  phase-space distribution  of Sgr debris are induced by the  LMC.  These
  perturbation are the  result of both the torque  exerted by the LMC
  on Sgr and the response of the MW to the LMC's gravitational pull.}
\label{fig:hs}
\end{figure*}

\section{N-body models of the phase-space distribution of Sgr-like tidal debris}
\label{sec:nbody_sgr}

Thus  far, we  have explored the  effects 
of  allowing the  MW's center  of mass to respond to perturbations from its satellites 
on the history of the Sgr dwarf galaxy's orbit.
However,  such analytic arguments are insufficient to explore the significance of such perturbations
on the  phase-space distribution of the  Sgr stellar stream. To explore this, 
we run  a new set of experiments, now based on  full $N$-body numerical simulations 
following the evolution of Sgr forward in time. We describe the models and methodology in 
Section~\ref{sec:nbody_models}, and present our results in Section~\ref{sec:nbody_res}.

\subsection{$N$-body models}
\label{sec:nbody_models}

\begin{table}
\begin{minipage}{90mm} \centering
  \caption{Summary of the set-up for the $N$-body  simulations
    analyzed in Section~\ref{sec:nbody_sgr}.}
\label{tab:N_model}
\begin{tabular}{@{}llllr} 
\hline 
\hline
{\bf Host} & & & \\ 
DM halo & & & $N_{\rm part} = 2.65 \times 10^{5}$ \\
Virial mass & $1 \times 10^{12}$ & & $[M_{\odot}]$ \\ 
Scale radius & $26.5$ & & [kpc] \\
Concentration & 9.86 & & \\
& & & \\
Stellar disk & & & $N_{\rm part} = 3 \times 10^{4} $ \\
Mass & $6.5 \times 10^{10}$ & & [$M_{\odot}$] \\
Scale length & 3.5 & & [kpc] \\
Scale height & 0.53 & & [kpc] \\
& & & \\
Stellar bulge & & & $N_{\rm part} = 5 \times 10^{3} $ \\
Mass & $1 \times 10^{10}$ & & [$M_{\odot}$] \\
Scale radius & 0.7 & & [kpc] \\
\hline
\hline
{\bf Sgr Satellites} & & & \\
DM halo & {\it Light}  & {\it Heavy} &  $N_{\rm part} = 2.65 \times 10^{4}$ \\
Virial mass & $0.32 \times 10^{11}$ & $1 \times 10^{11}$  &[$M_{\odot}$] \\ 
Scale radius & $4.9$ & $6.5$ &  [kpc] \\
& & & \\
Stellar spheroid & & &  $N_{\rm part} = 5 \times 10^{4} $ \\
Mass & $6.4 \times 10^{8}$ & $6.4 \times 10^{8}$ & [$M_{\odot}$] \\
Scale radius & $0.85$ & $0.85$ & [kpc] \\
\hline
\hline
{\bf LMC Satellite} & & & \\
Single spheroid & & &  $N_{\rm part} = 2 \times 10^{4} $ \\
Mass & $1.8 \times 10^{11}$ &  & [$M_{\odot}$] \\
Scale radius & $20$ & & [kpc] \\
& & & \\
\end{tabular}
\end{minipage}
\end{table}

\subsubsection{Methodology}

The  $N$-body systems are  evolved using  GADGET-2.0 \citep{2005MNRAS.364.1105S}, a
well-documented, massively parallel  Tree-SPH code.  To  construct self-consistent 
stable models  of the MW, the LMC,  and Sgr, we follow
the  procedure  described   by  \citet{2008MNRAS.391.1806V}.  In  the following sections 
we  describe the  main properties of  each galactic
model. In general, the force softening is chosen  to be a tenth  of the mean
interparticle  distance  of each  system,  calculated using  particles
located within a distance of ten scale length radii. In particular,
when dealing with Plummer models we compute our softening lengths as
described by \citet{2000MNRAS.314..475A}. 

Keeping a MW model fixed in a $N$-body
simulation is significantly more challenging than in simulations with analytic, 
smooth galactic models. Having illustrated in Section~\ref{sec:ana_sgr} that perturbations on the orbit 
and debris of Sgr when fixing the MW are quite significant, in what follows we will only consider 
free MW models with and without the presence of a massive LMC. Note that, due to the relatively low mass of 
Sgr, the pertubative effects associated with the phase-space 
displacement of the MW center of mass in previous studies that have only considered the interaction 
between the MW and Sgr  (i.e. disregarding the LMC) are negligible. 

\subsubsection{Galactic models}
\label{sec:pot_gal_ndist}

We model the MW-like host as a self-consistent three-component
system containing a NFW dark matter halo, an exponential stellar disk,
and a  central bulge following a Hernquist profile.  The  dark matter
halo has  a total mass of  $M_{\rm vir} =  10^{12}~M_{\odot}$, a scale
radius  $r_{\rm  s}  =  26.47$  kpc,  and  is  initially  adiabatically
contracted to model its response to the formation of a stellar disk in its
central  region  \citep{1986ApJ...301...27B, 1998MNRAS.295..319M}.   The
exponential disk  has a total  mass of $6.5  \times 10^{10}~M_{\odot}$
and a scale  length and height of  3.5 and 0.53  kpc, respectively. For
the bulge we assume a mass of $10^{10}~M_{\odot}$ and a scale radius
of  0.7  kpc. As previously discussed, the circular  velocity
profile  takes a value of
$\sim 239$ km s$^{-1}$ at $\sim 8.29$ kpc from the galactic center.

We  use a  Plummer distribution  to model  the LMC.   For this  set of
numerical experiments we consider a profile with  a total mass of $M_{\rm LMC} =
1.8 \times  10^{11}~M_{\odot}$ and a  scale radius $r_{\rm LMC}  = 20$
kpc.  Since the mass spreads out to infinity in Plummer models, the
density profile is initially truncated at the radius that encloses $95\%$ of 
the LMC's total  mass.    

Based on  \citet{2011Natur.477..301P}, the Sgr 
progenitor is self-consistently initialized with a NFW dark matter
halo and a spheroidal stellar component that follows a Hernquist
profile. Two different dark matter mass models are considered:
\begin{enumerate}
\item a ``Light'' model with a DM halo mass of $10^{10.5}$ M$_{\odot}$;
\item a ``Heavy'' model with a DM halo mass of  $10^{11}$ M$_{\odot}$.    
\end{enumerate}
The  stellar
components in both models have a total mass  of $6.4 \times 10^{8}$ M$_{\odot}$ and a
scale radius of  $0.85$ kpc \citep{2012MNRAS.422..207N}.  As discussed
in  Section~\ref{sec:pot_gal_smooth}  (see  also  P11),  the  Sgr-like
satellites are launched at a Galactocentric distance of 80 kpc 
from the Galactic centre in  the plane of the MW disk, traveling
vertically  at 80  km s$^{-1}$  towards the  North Galactic Pole. 
To  account  for  the  mass  loss that  would  have  occurred  between the crossing of the MW's
virial radius and this ``initial'' location, the Sgr progenitor
NFW  mass  profiles  are  initially  truncated  at  the  corresponding
instantaneous  Jacobi radius, $r_{\rm J} \approx 30$ and 23 kpc for the 
Heavy and Light Sgr, respectively.

Table~\ref{tab:N_model} summarizes the values of all of the parameters 
that describe our $N$-body models.

\begin{figure*}
\centering
\includegraphics[width=85mm,clip]{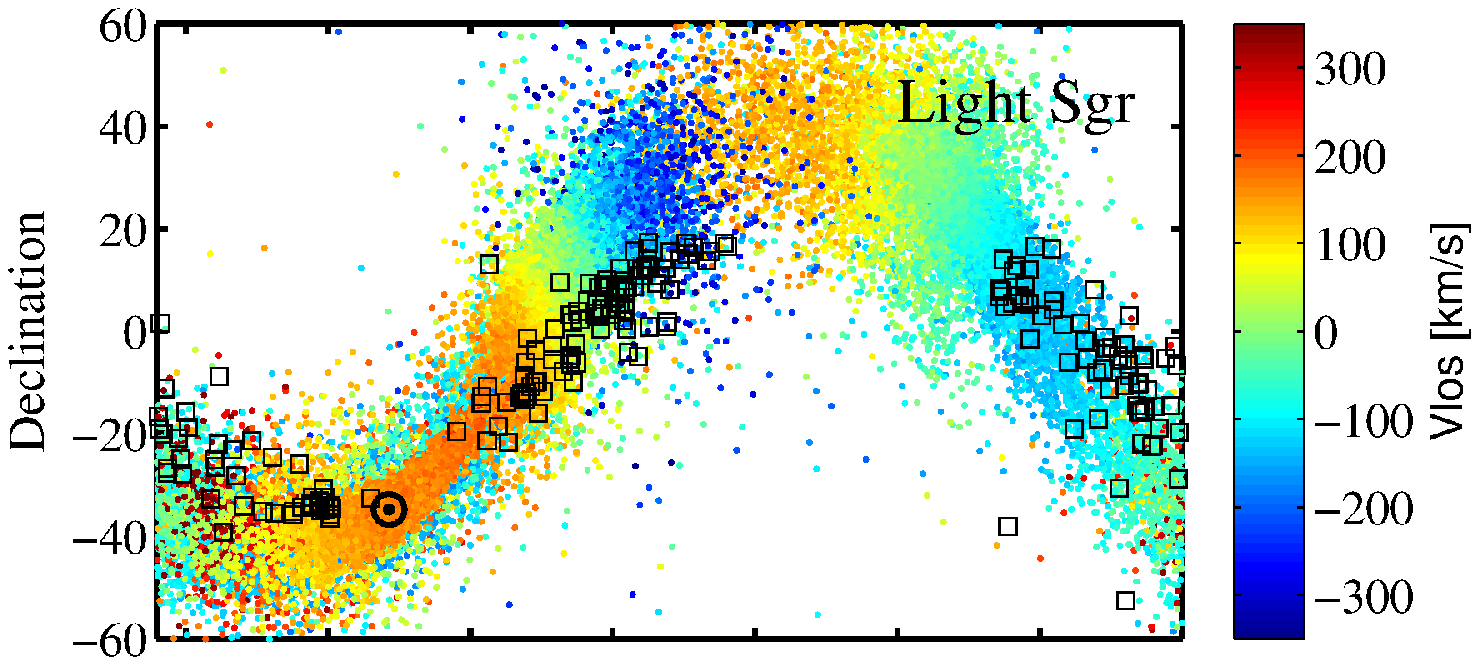}
\includegraphics[width=85mm,clip]{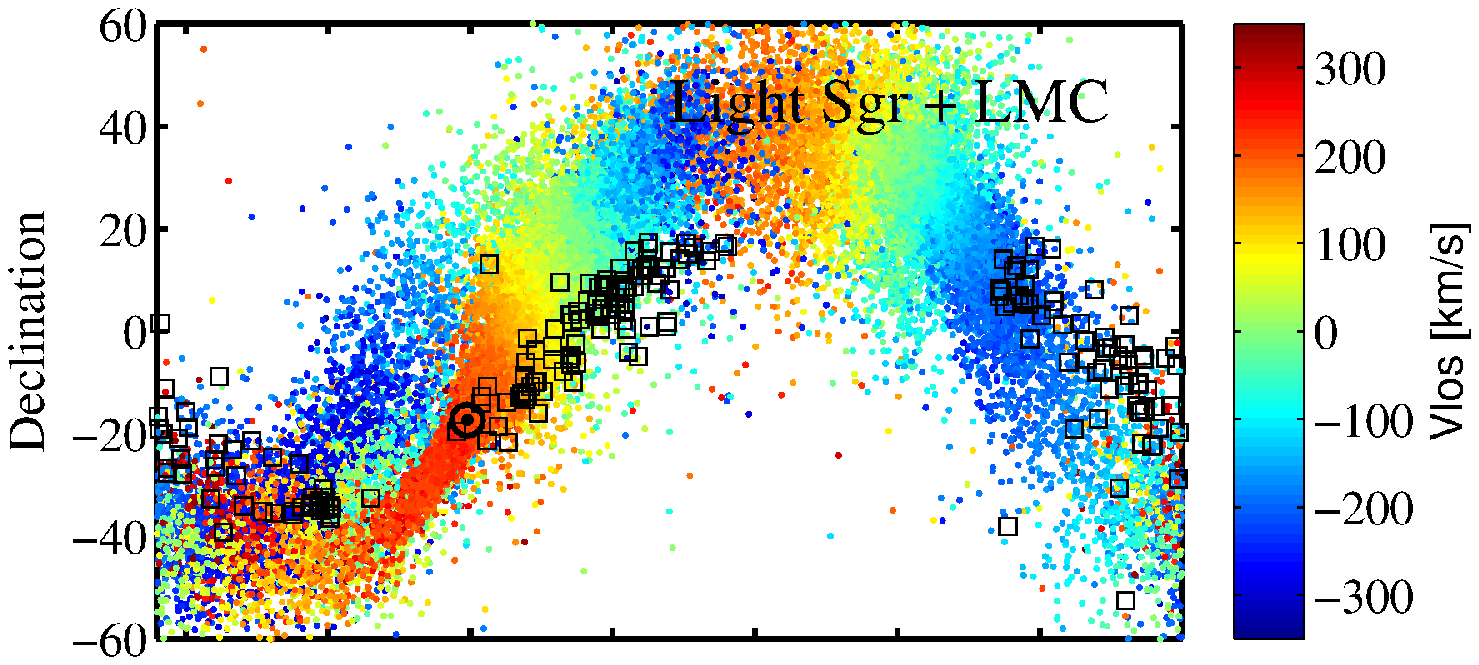} \\
\includegraphics[width=85mm,clip]{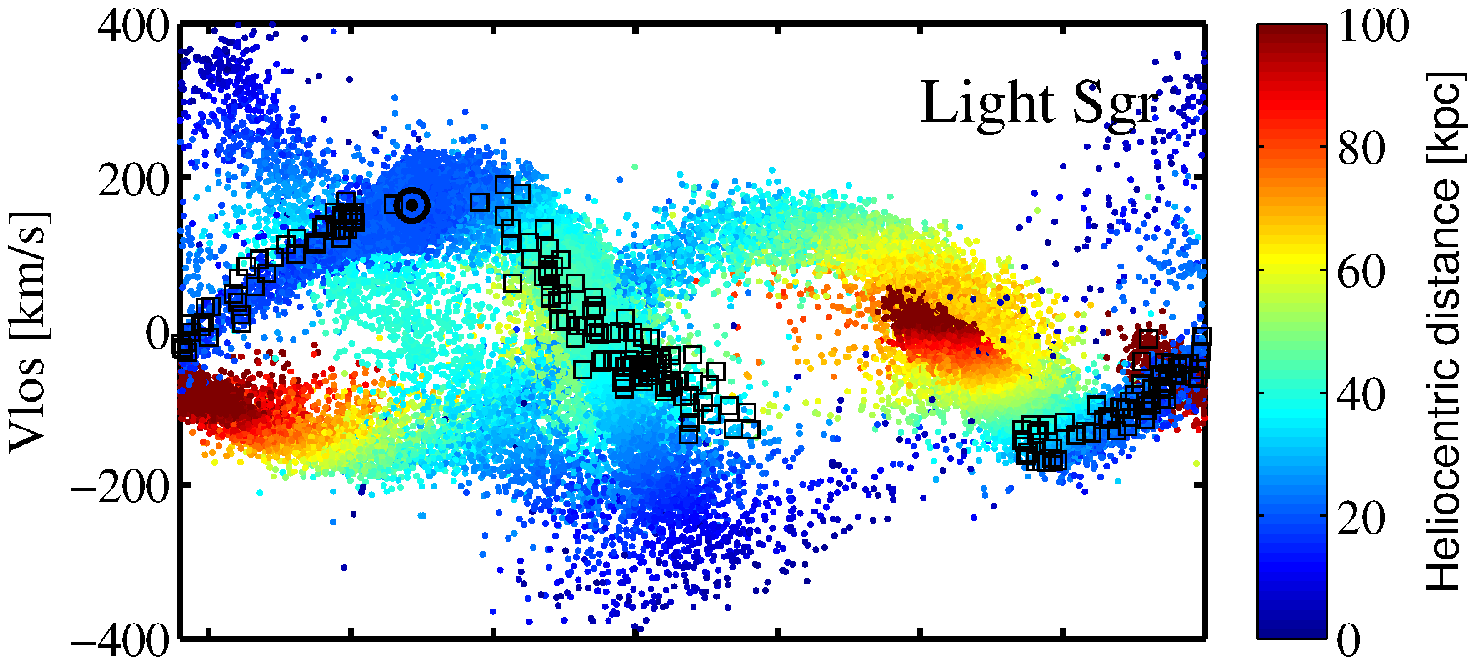}
\includegraphics[width=85mm,clip]{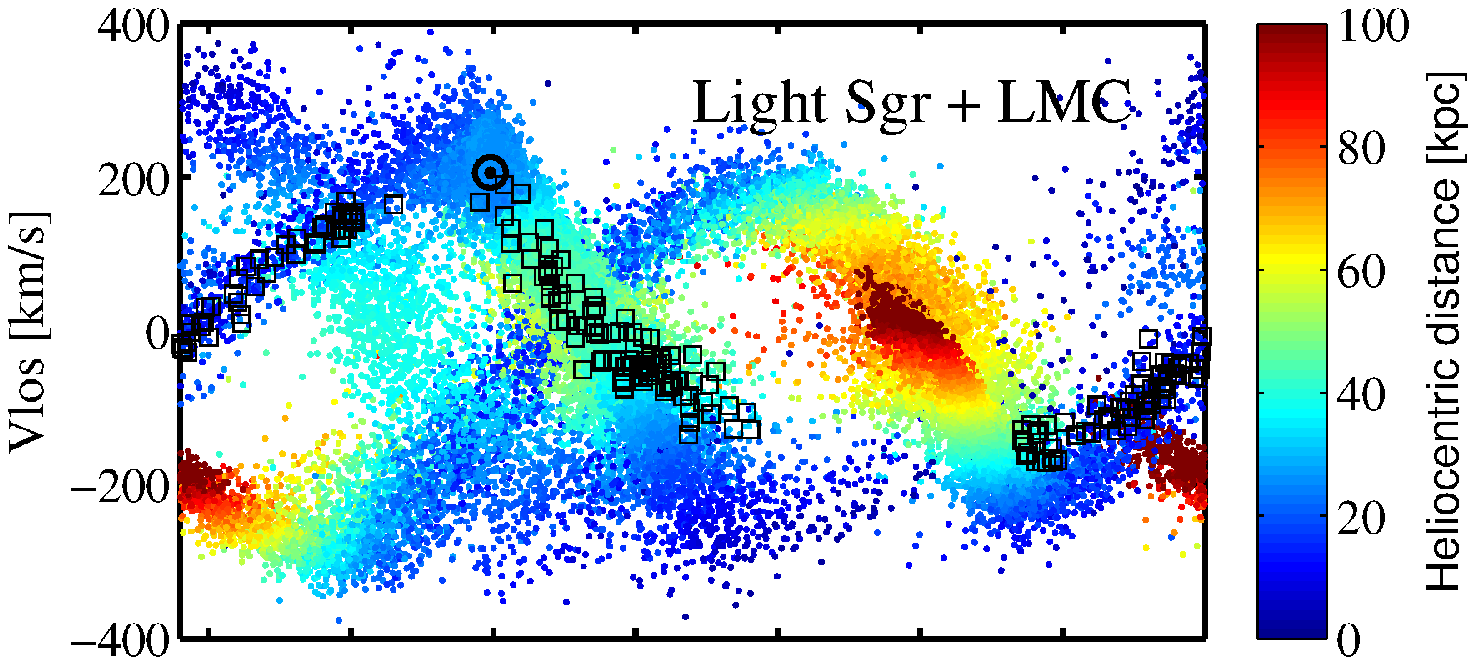} \\
\includegraphics[width=85mm,clip]{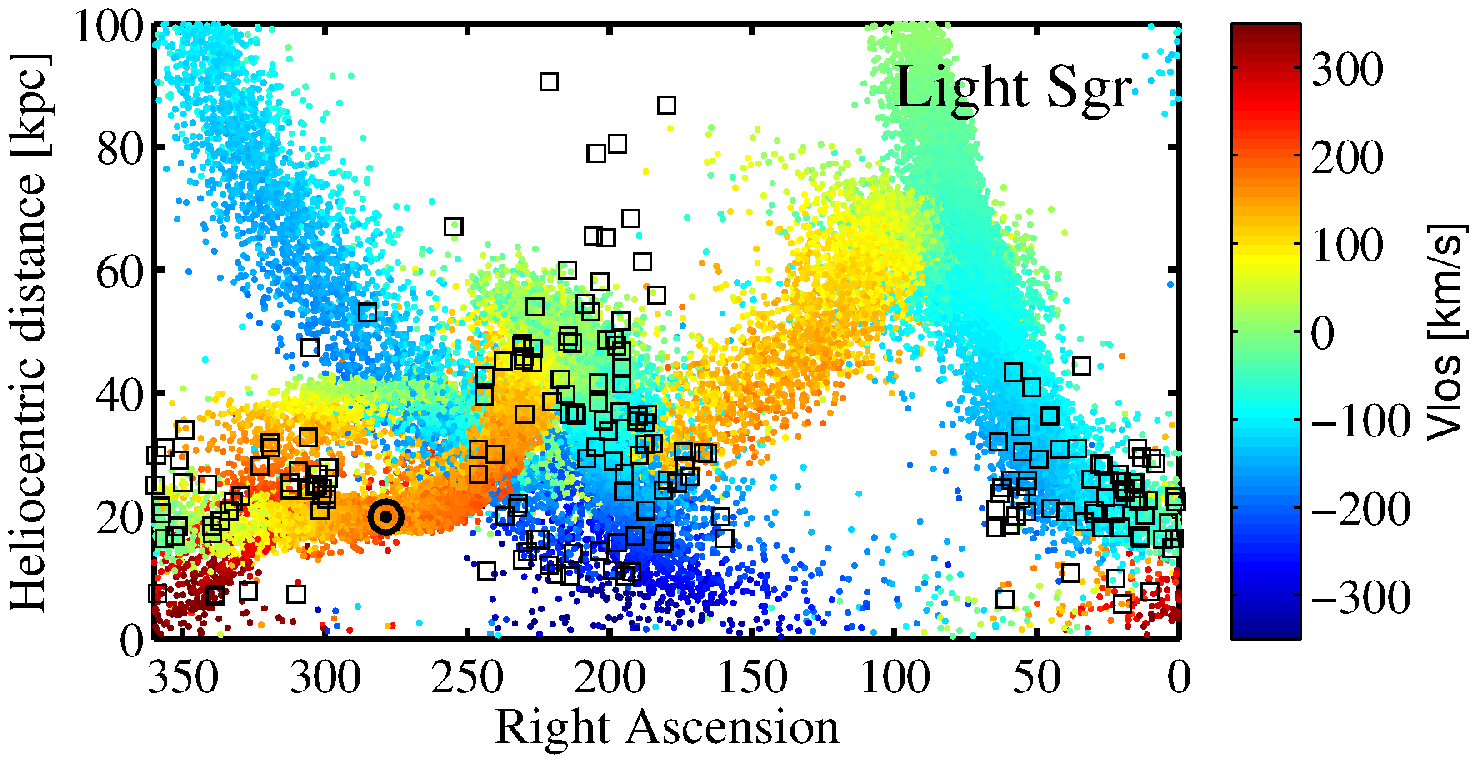}
\includegraphics[width=85mm,clip]{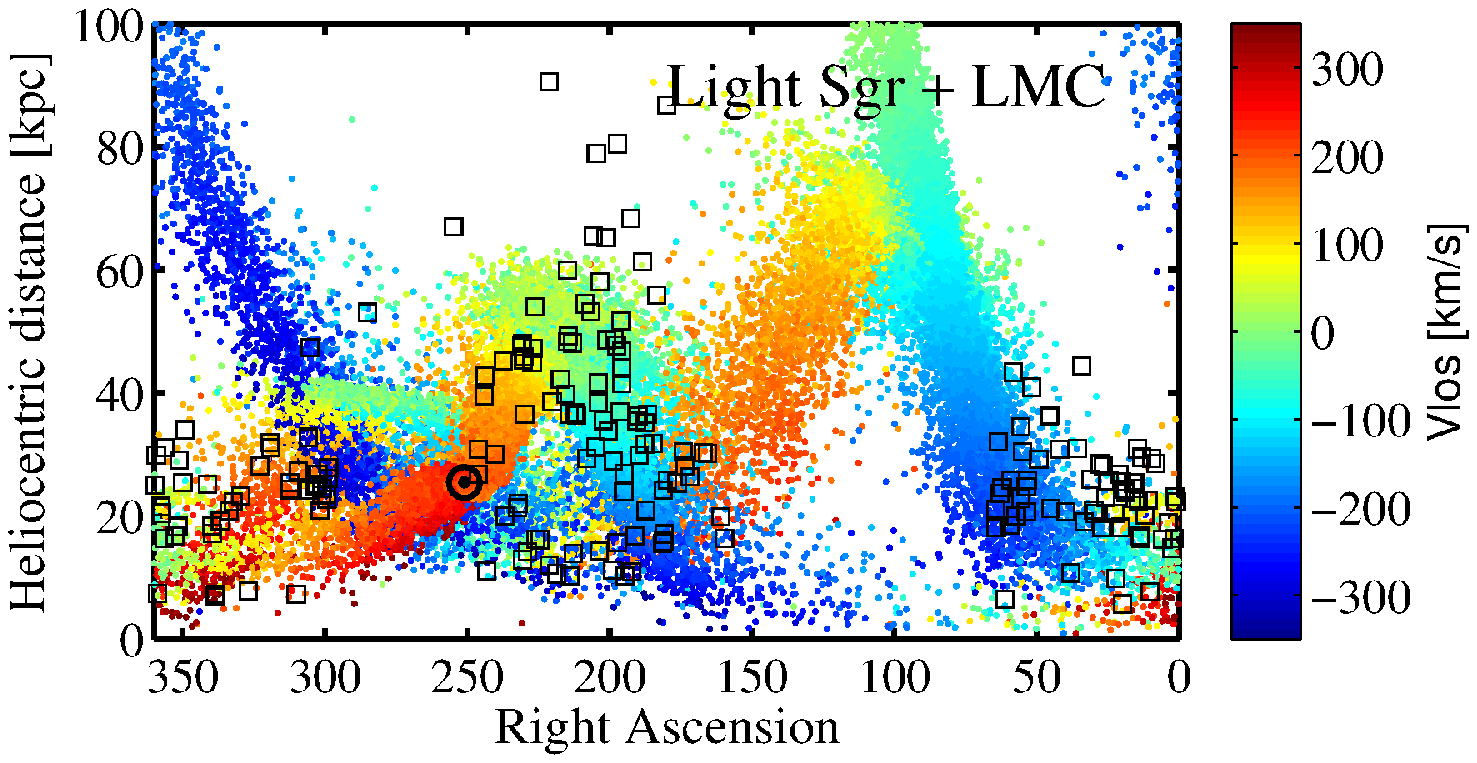}
\\
\caption{As in Figure~\ref{fig:hs}, for  the simulation with the Light
  Sgr  model.   Note  again  the  significant   perturbations  to  the
  phase-space distribution of Sgr debris induced by the LMC.}
\label{fig:ls}
\end{figure*}

\subsection{An $N$-body treatment of the MW + LMC + Sgr system}
\label{sec:nbody_res}

The $N$-body models discussed in this section will build on the previously-discussed
analytic models to explore the impact of LMC perturbations in a first infall scenario on the Sgr tidal debris.
To this end, we compare simulations in which the LMC is included against others in which it is not.
In what follows,   all   simulations  include   fully   self-consistent
three-component MW models that  are allowed to respond  to the
gravitational  pull   of  any   external  source. As   described  in
Section~\ref{sec:pot_gal_ndist}  we  consider  Heavy  and Light  Sgr
models, both  self-consistently initialized with spherical baryonic and dark
matter components. The Heavy and the Light Sgr
are followed for $\sim 2.1$ and $2.6$ Gyr, respectively, at which time
they reach a heliocentric distance  of approximately 20 kpc. For more
details on the setup of these simulations, we refer the reader to P11.

In the  left panels of  Figures~\ref{fig:hs} and \ref{fig:ls}  we show
the distribution of the Sgr debris at present-day in different projections
of phase-space  in the Heavy and Light  Sgr simulations, respectively.
In this simulation the LMC is not included, but the MW can respond to the presence 
of Sgr.
The   black    squares   show   data   from    2MASS   M-giant   stars
\citep{2004AJ....128..245M}.   As  discussed   by  P11,  although  the
simulated Sgr debris  distributions do not precisely match  all of the
observed characteristics, they do  produce a reasonable fit. Note that
small differences in these distributions  of debris and those shown in
Figure S3  of P11 are probably due to the  slightly different MW model
used in this  work. 

The right panels show  the same distributions, but now
including in the simulations  a first-infall LMC model
with a total  mass $M_{\rm LMC} = 1.8  \times 10^{11}~M_{\odot}$.  The
initial  conditions for  the LMC  models were  obtained  by backward
time integration from  its present-day location in  a ``free'' MW
scenario (see Section~\ref{sec:lmc}) until 2.1 or 2.6 Gyrs ago, 
depending on the Sgr model. Note that, due to the slight overestimation of the role of
dynamical friction in the analytic calculations (see discussion in Section~\ref{sec:mw_lmc_ana}), 
the LMC initial conditions (ICs) were iteratively calibrated by  comparing 
the resulting LMC $N$-body orbits with their analytic counterpart. 
The goal of this exercise was to obtain a set of ICs 
for the $N$-body simulations that, at present day, yields the correct phase-space coordinates.    
 The Sgr dwarf galaxy
is launched with the same  initial conditions as in  the ``LMC-less''
simulations.  Clearly, the addition of the LMC results in significant
perturbations in  the phase-space distribution of Sgr  debris.

The top panels of Figures~\ref{fig:hs} and \ref{fig:ls}
show the simulated present-day  Sgr stream and the remnant core projected
in right ascension (RA) and declination. A comparison between simulations
with and without the LMC model reveals an interesting feature at RA
$\approx 300^{\circ}$. When the LMC is accounted 
for,  tidal debris that  otherwise would overlap  when projected on the
sky are split into two distinguishable arms. This is true for both Sgr models,
suggesting that tidal material could also be deposited in regions of the sky that 
are not delineated by the present-day Sgr orbital plane.  
Note that the two arms  show both opposite
heliocentric   distance   gradients as a function of RA  (see   bottom   right  panel   in
Figures~\ref{fig:hs} and \ref{fig:ls}) and opposite line-of-sight velocities.   

\begin{figure}
\centering
\includegraphics[width=80mm,clip]{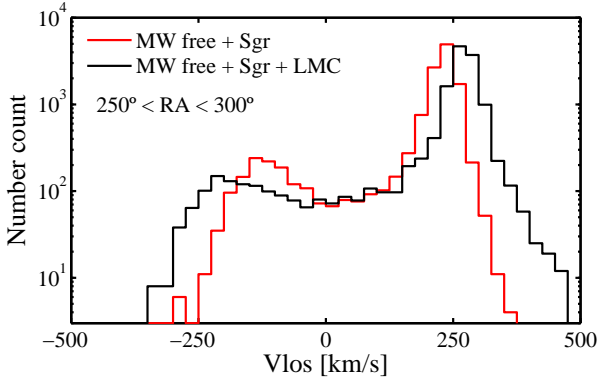}
\caption{Distribution of lines-of-sight-velocities, $V_{\rm  los}$, of 
Heavy Sgr star particles located within 
$250^{\circ} < {\rm RA} < 300^{\circ}$. The red line shows
the results obtained in a self-consistent $N$-body simulation of the interaction
between the MW and Sgr. The black line shows
the results obtained when a model of the LMC is added to the simulation.}
\label{fig:vlos_comp}
\end{figure}

The middle panels of Figures~\ref{fig:hs} and \ref{fig:ls}
show   the  projected   Sgr   distribution  in   RA  versus  Galactic
standard-of-rest  line of  sight  velocity ($V_{\rm los}$) space.   A
quick  comparison  between  the  left  and right  panel  reveals  very
significant changes in the distribution of $V_{\rm los}$.  In general,
adding  the LMC  results in  a much  broader distribution  at  all RA. 
This can be more clearly seen on Figure~\ref{fig:vlos_comp} where, as an 
example, we show the $V_{\rm los}$ distributions of Heavy Sgr star particles located 
within $250^{\circ} < {\rm RA} < 300^{\circ}$.
Recall   that,   as    shown   in   Figure~\ref{fig:diff_lmc},   these
perturbations are  not just the result  of the LMC torque  on Sgr, but
are also due  to the self-consistent response of the  MW to the
LMC's gravitational  pull. The  phase-space distribution of the Sgr debris
obtained  when the LMC  is included  in the  simulation results  in a
worse fit to the  \citet{2004AJ....128..245M} data.  However, 
in this work we have not attempted to find a set of initial conditions
that  could fit  the Sgr  debris in  a scenario  in which  the  LMC is
included. Starting with different initial conditions for the Sgr orbit or 
a lower LMC mass could plausibly bring the velocities into better agreement.   
Instead,  our goal  is  simply  to  explore what perturbations
are induced and whether they are significant.

Perturbations to  the Sgr
debris  phase-space distribution can  also be  observed in  the bottom
panels of Figures~\ref{fig:hs} and \ref{fig:ls},  where we  show  the projection onto  RA versus  heliocentric
distance  space. It is clear that the addition of the LMC resulted in a 
significant spatial redistribution of Sgr debris. 
Note that, independent of whether the LMC is included or not,
 star  particles in the  leading and trailing arm can reach distances of  $\sim
50$ kpc (at RA $\approx  240^{\circ}$) and $\sim 100$  kpc  (RA  $\approx
80^{\circ}$),\footnote{Similar results are obtained in a Galactocentric
reference  frame} respectively (also, see Figure S4 from P11). 
These different distances  are similar  to the leading 
and trailing tail's apocentric distances of the Sgr stream, as traced by  B14.
They find $R_{\rm lead}^{\rm apo} \approx 48$ kpc and 
$R_{\rm trail}^{\rm apo} \approx 102$ kpc, respectively. The different  
apocentric distances reached by the star
particles in our simulations in the leading and trailing arms are merely a consequence
of       considering      a      self-gravitating       Sgr      model
\citep[see][]{2007MNRAS.381..987C, 2014MNRAS.445.3788G}.

\begin{figure}
\centering
\includegraphics[width=80mm,clip]{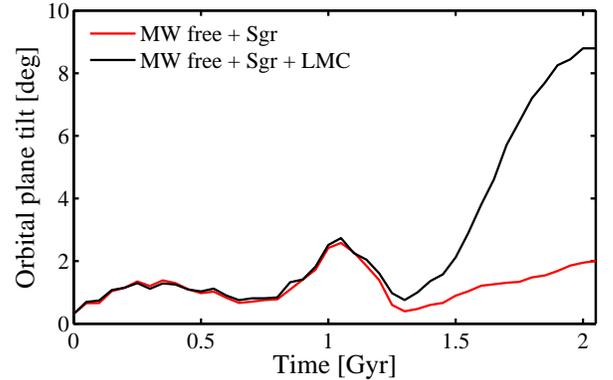}
\caption{Time  evolution  of   the  Heavy  Sgr  orbital  angular  momentum
orientation with respect to its initial direction. The red line shows
the results obtained in a self-consistent $N$-body simulation of the interaction
between the MW and the Sgr dwarf galaxy. The black line shows
the results obtained when a model of the LMC is self-consistently
added to the simulation. }
\label{fig:plane_tilt}
\end{figure}

It is also interesting to explore whether the self-consistent addition
of the LMC  could at least partially explain  the $\approx 10^{\circ}$
difference  between  the mean  orbital  poles  of  the great  circles
associated    with    the    debris    leading    and    trailing    Sgr
\citep{2005ApJ...619..800J}.   In Figure~\ref{fig:plane_tilt}  we show
the  time  evolution  of   the  Heavy  Sgr  orbital  angular  momentum
orientation, $\hat{L}$.  Since  Sgr is launched in the  X-Z plane, its
angular momentum initially points in the $\hat{Y}$ direction.  The red
line  shows the  angular  displacement of  $\hat{L}$  with respect  to
$\hat{Y}$ in the LMC-less scenario.  As expected from a polar orbit in
an axisymmetric  potential, the  orientation of the  angular momentum
remains nearly  constant and  close to $0^{\circ}$  at all  times. The
black   line  shows  the   result  obtained   after  adding   the  LMC
model. Clearly, as  the LMC approaches the MW, the Sgr orbital
plane  starts to  tilt with  respect to  its initial  orientation. This
tilting  takes place during  the last  0.5 Gyr  of evolution,  in good
agreement  with  the  results  shown  in  Figure~\ref{fig:mw_com}.  At
present-day, the angular displacement of $\hat{L}$ is of approximately
$9^{\circ}$,   similar to  the   value  reported   by
\citet{2005ApJ...619..800J}.

\section{Discussion and conclusions}
\label{sec:conc}

In  this  work we  have  performed and  analyzed  a  set of  numerical
simulations  using smooth  and  $N$-body gravitational potentials. Our  goal was  to
explore whether the approximation of an inertial Galactocentric reference frame holds 
in the presence of a relatively massive LMC that is experiencing its
first infall towards the MW. In a nutshell, if the LMC currently has a total mass
of at least $ 5 \times 10^{10}~M_{\odot}$, the answer is likely to
be no. 

To  arrive to this conclusion, we have focused
our efforts  on two possible  situations where  artificially fixing
the MW center  of mass could  have a significant  effect.  Our
first obvious choice  was to explore the implications on the
orbit calculations of the LMC about the MW. Our results clearly
show that the LMC's orbital period and apocentric distance are significantly 
shortened if we allow the MW  to  react to the LMC's  gravitational pull.  
As the mass of the LMC becomes larger, and thus more comparable to the MW mass 
enclosed within the LMC's location, the two-body interaction becomes more relevant. 
Thus, the more  massive the  LMC, the  larger the  changes on its orbital periods.
The change in the inferred orbital properties of our 
LMC-like models suggest that, even though a first-infall 
is still a very plausible scenario, the limiting LMC-MW mass combinations 
that could host a first-infalling LMC are noticeably affected; 
it raises the required minimum LMC mass and disfavors MW models with $M_{\rm vir} \geq 1.5 \times 10^{12}~M_{\odot}$.
A detailed dynamical analysis, including  $N$-body models that can 
naturally account for the two body interaction, the LMC's tidal mass loss and dynamical friction, 
and a model for the time evolution of the MW potential, would be required to robustly characterize 
the orbital history of the different LMC mass models. This is beyond the scope of the work presented in this paper.

We have  also characterized how the MW itself responds to the gravitational pull of the
LMC. We find that significant changes in both the position and velocity of
the MW center of mass takes  place only during the  last $0.3-0.5$~Gyr 
of evolution. It is around this time  when both the MW mass
enclosed within  the LMC  Galactocentric distance becomes  comparable to
the mass of the LMC itself and the distance between both galaxies becomes short enough. 
For example, for a LMC model with a total mass $M_{\rm LMC}
= 1.8 \times 10^{11}~M_{\odot}$,  the orbital barycenter is located at
$\approx 14$ kpc from the MW  center of mass at the present day.  For this
LMC model,  the MW was displaced by $\approx  30$ kpc and
its  velocity changed by  $\approx 75$  km/s in  this very short
amount of time. Note that similar results were obtained in simulations of the collision 
between Andromeda and its satellite galaxy M32 \citep{marion}. 

Due to  the extended  nature of  the MW  stellar halo,  not all
stars  are accelerated at  the same  rate  by a massive satellite.  
It is thus likely  that this
differential  acceleration  will  have  important effects  on  the
observable properties  of extended  stellar streams. For  example, the
distribution of Sgr  debris, which covers a radial  extension of $\sim
100$  kpc, could  be significantly  perturbed due  to  the phase-space
displacement of  the Milky  Way center of  mass, in addition to the perturbations 
associated with the LMC torque (e.g. VCH13).  To explore  this, we
integrated Sgr-like orbits in MW potentials that are allowed to
freely react  to the gravitational pull  of the LMC. We  would like to
stress  that in  this work  we have  not attempted  to find  a  set of
ICs for  the Sgr  progenitor  that could  fit the  Sgr
debris in a scenario in which  the LMC is included. The complexity behind the search 
for best-fitting ICs in a three-body problem scenario using fully self-consistent 
models is beyond the scope of this work. Instead, our goal
is simply  to explore whether or not such perturbations are significant.

Our  analysis showed  that,  indeed,  the  presence of  the  LMC
introduced  significant  perturbations  on  Sgr-like orbits  and  their
associated distribution of debris.  We have confirmed previous results
presented by VCH13, where they showed that the torque on Sgr exerted by
the LMC can introduce  non-negligible perturbations on its orbit and
distribution of debris. However,  we find that the differences between
the Sgr-like orbits  obtained in ``free'' and a  ``fixed'' MW +
LMC models are even larger that those obtained in ``free'' models with
or without the LMC. Furthermore,  we have shown that this perturbation
is significant even in the scenario where the LMC is undergoing its
first  pericenter  passage.   Attempts  to reproduce  the  Sgr  stream
without a model  for the LMC perturbation will  thus force searches for
the  best-fitting  parameters   that  characterize  the   MW gravitational potential to 
artificially adjust in order to account for this  perturbation. 

An  example  is  the  discrepancy  discussed  by
\citet[][B14]{2014MNRAS.437..116B} in the angular distance between the
inferred apocenters of the Sgr leading and trailing arms. Observations
suggest that this  angular distance is smaller than  what is predicted
in a fixed  logarithmic  potential.    Our  analysis  showed  that  the
differences in  the angular distances between the  last two Sgr's orbital apocenters
could at least be partially accounted for with both a free MW model 
and the inclusion of the LMC. 

Another  example  is the  $\approx
10^{\circ}$ difference  between the mean  orbital poles of  the great
circles associated with the debris  leading and trailing the  Sgr core,
reported  by \citet{2005ApJ...619..800J}. We  find in  our simulations
that,  due to  the  gravitational pull  exerted  by the  LMC, the  Sgr
orbital  plane  tilts  with  respect  to its  initial  orientation  by
$\approx 9^{\circ}$  during the last  0.5 Gyr of evolution.  These are
just two  examples of peculiar  characteristics of the Sgr debris
that could be  naturally and,  at least, partially accounted for if 
a  fully self-consistent model of the MW + LMC + Sgr interaction is considered. 

Interestingly, these results were obtained without the need for a prolate/oblate model of the Galactic DM halo.
To accurately quantify the significance of these perturbations, fully self-consistent models 
of the MW + LMC + Sgr interactive system are required. Note that, for each combination of galactic 
models, a specially tailored set of initial orbital conditions for the LMC and Sgr will be 
required. We defer this analysis to a follow-up work.

The orbit of the LMC about the MW and the orbital history and phase-space distribution of Sgr 
debris are just two examples where perturbations induced by the MW + LMC interaction could be 
significant. The inferred orbital properties of other MW dwarfs, such as Carina, Fornax, Sculptor and 
Ursa Minor, obtained using present-day phase-space coordinates, could also be affected by such 
interaction if the LMC is massive enough \citep[e.g.][]{2011A&A...525A..99P,2011MNRAS.416.1401A}.
Furthermore, using HST proper-motion measurements, \citet{2012ApJ...753....8V} 
estimated a radial velocity of M31 with respect to the MW of $V_{\rm rad,M31} = -109.3 \pm 4.4$ km/s
, and a tangential velocity $V_{\rm tan,M31} = 17.0$ km/s, with $1\sigma$ confidence region 
$V_{\rm tan,M31} \leq 34.3$ km/s. We have shown  that, if the LMC is as massive as 
$1.8 \times 10^{11}~M_{\odot}$, the velocity of the MW center of mass could have changed by as much as 75 
km/s in less than 0.5 Gyr. Decomposing this velocity into a tangential and radial components toward M31 
yields $V_{\rm rad,MW} \approx 37$ km/s and $V_{\rm tan,MW} \approx 66$ km/s. This suggests that estimates of 
the Local Group mass based on timing arguments could be affected by such a two-body interaction. In addition,
a significant fraction of the present-day relative velocity of M31 with respect to the Galactic centre 
could be associated to the temporary Galactic displacement about its orbital barycenter, thus affecting the 
projected evolution of the MW + M31 system.

We are on the verge of the so-called {\it Gaia} era.  In addition to the
very accurate  phase-space catalogs that  we are already  mining, {\it
  Gaia} is starting to collect phase-space information for many millions of
stars. The  high-quality data  that will soon  become available clearly
calls for the development of models that are as detailed as possible, and which include all known 
sources of significant interactions. The results presented in this work suggest that,
if the LMC is as massive as suggested by recent studies, to properly interpret this data 
it is essential to consider in the analyses self-consistent MW + LMC models that are allowed 
to freely react to their mutual gravitational interactions. 

\acknowledgments FAG and BWO are supported through the NSF
Office of Cyberinfrastructure by grant PHY-0941373 and by the 
Michigan State University Institute for
Cyber-Enabled Research (iCER).  BWO was supported in part by NSF grant PHY
08-22648: Physics Frontiers Center/Joint Institute for Nuclear
Astrophysics (JINA). DDC is supported by the Universidad Nacional de 
La Plata, Argentina, and the Instituto de Astrof{\'i}sica de La Plata, 
UNLP-Conicet, Argentina. G.B. acknowledges support from
NASA through Hubble Fellowship grant HST-HF-51284.01-A.

\bibliographystyle{apj}
\bibliography{apj-jour,mw_lmc_sgr}

\label{lastpage}
\end{document}